\documentclass[aps,prb,reprint,twocolumn,floatfix,superscriptaddress,amsmath,amssymb,amsfonts]{revtex4-2}
\AtBeginDocument{\usepackage{booktabs}}               
\makeatletter

\usepackage{graphicx}
\usepackage{dcolumn}
\usepackage{bm}
\usepackage{xcolor}
\usepackage{multirow}
\usepackage[percent]{overpic}
\usepackage[newcommand]{ragged2e}
\usepackage{rotating}
\usepackage{tikz}
\usepackage{gensymb}
\usepackage{mathtools}
\usepackage{mhchem}

\usepackage{caption}
\DeclareCaptionJustification{justified}{\justifying}
\usepackage{subcaption}

\usepackage[colorlinks=true, urlcolor=blue, linkcolor=blue, citecolor=blue, pdftex]{hyperref}
\usepackage[mathlines]{lineno}

 \g@addto@macro\bfseries{\boldmath}
\makeatother

\usepackage{float}

\newcommand{\3}{\hspace*{-3pt}}

\begin{document}

\preprint{APS/123-QED}

\title{Pseudo-fermion functional renormalization group with magnetic fields}

\author{Vincent Noculak}
\affiliation{Dahlem Center for Complex Quantum Systems and Fachbereich Physik, Freie Universit\"at Berlin, Arnimallee 14, 14195 Berlin, Germany}
\affiliation{Helmholtz-Zentrum Berlin f\"ur Materialien und Energie, Hahn-Meitner-Platz 1, 14109 Berlin, Germany}
\author{Johannes Reuther}
\affiliation{Dahlem Center for Complex Quantum Systems and Fachbereich Physik, Freie Universit\"at Berlin, Arnimallee 14, 14195 Berlin, Germany}
\affiliation{Helmholtz-Zentrum Berlin f\"ur Materialien und Energie, Hahn-Meitner-Platz 1, 14109 Berlin, Germany}
\affiliation{Department of Physics and Quantum Center for Diamond and Emergent Materials (QuCenDiEM), Indian Institute of Technology Madras, Chennai 600036, India}

\date{\today}

\begin{abstract}
The pseudo-fermion functional renormalization group is generalized to treat spin Hamiltonians with finite magnetic fields, enabling its application to arbitrary spin lattice models with linear and bilinear terms in the spin operators. We discuss in detail an efficient numerical implementation of this approach making use of the system's symmetries. Particularly, we demonstrate that the inclusion of small symmetry breaking magnetic seed fields regularizes divergences of the susceptibility at magnetic phase transitions. This allows the investigation of spin models within magnetically ordered phases at $T=0$ in the physical limit of vanishing renormalization group parameter $\Lambda$. We explore the capabilities and limitations of this method extension for Heisenberg models on the square, honeycomb and triangular lattices. While the zero-field magnetizations of these systems are systematically overestimated, the types of magnetic orders are correctly captured, even if the local orientations of the seed field are chosen differently than the spin orientations of the realized magnetic order. Furthermore, the magnetization curve of the square lattice Heisenberg antiferromagnet shows good agreement with error controlled methods. In the future, the inclusion of magnetic fields in the pseudo-fermion functional renormalization group, which is also possible in three dimensional spin systems, will enable a variety of additional interesting applications such as the investigation of magnetization plateaus.
\end{abstract}

\maketitle

\section{\label{sec:introduction}Introduction}

The ever growing landscape of quantum materials inspires new research directions, but poses significant challenges to a theoretical description, particularly, when strong electron correlations are involved. However, the hard shells of associated quantum many-body problems may contain soft cores with fascinating emergent phenomena and effective theories that are worth exploring. A prominent example in the realm of spin systems are quantum spin liquids characterized by effective gauge theories, strong entanglement, and exotic quasiparticles~\cite{Broholm20,Knolle19,Savary17}.

Cracking the hard shell of a quantum many-body problem usually requires sophisticated numerical techniques.
Decades of method development in the field of frustrated magnetism have produced a wealth of powerful numerical approaches that have led to impressive successes, but have also revealed shortcomings that each method is plagued with. For example, exact diagonalization only allows for the treatment of small systems. Tensor network approaches, such as DMRG, effectively reduce the considered Hilbert space such that larger systems become accessible~\cite{orus_tensor_2019,Cirac21}. However, as a result of the entanglement scaling, tensor network techniques are often seriously challenged by dimensions larger than one. Quantum Monte Carlo (qMC) approaches are, in principle, strictly error controlled, but their applicability is usually limited to unfrustrated interactions due to the infamous sign problem~\cite{VONDERLINDEN92}. The pseudo-fermion functional renormalization group (PFFRG) considered here represents an alternative approach to numerically investigate systems of frustrated magnetism~\cite{Mueller23}. Being complementary to the aforementioned techniques, it comes along with its own strengths and weaknesses.

The PFFRG uses a mapping of $S=\frac{1}{2}$ spins onto pseudo-fermions \cite{Abrikosov65,Affleck88} enabling the application of the functional renormalization group (FRG) formalism to the resulting fermionic Hamiltonian. The parent FRG approach relies on the introduction of a regularization parameter (usually called $\Lambda$) in the free single-particle Green function, allowing for the derivation of coupled differential equations for the $\Lambda$-dependent multi-particle vertex functions~\cite{Metzner12}. 

Since its initial formulation in 2010~\cite{Reuther10}, the PFFRG has been steadily improved~ \cite{Reuther11,Hering17,Buessen19,Gresista22,Schneider23, Mueller23}. First, it was applied to $S=\frac{1}{2}$ Heisenberg models at zero temperature~\cite{Reuther10}. Later it was shown that a generalization to models with anisotropic spin interactions is possible with only a moderate increase in complexity~ \cite{Hering17,Buessen19}.
Higher spin-$S$ models are treatable as well and it has been shown that the method becomes identical to the Luttinger-Tisza approach~\cite{Luttinger46,Luttinger51} in the classical large $S$ limit \cite{Baez17}. Similarly, a generalization from SU(2) to SU($N$) spins and an extrapolation to the large-$N$ limit can be performed~\cite{Buessen18}. While the PFFRG was initially proposed at zero temperature, a recent implementation of the Popov-Fedotov trick~\cite{Schneider23} or the usage of a Majorana spin representation~\cite{Niggemann21,Niggemann22} also allow for finite temperature applications.

Flexibility is a key feature of the PFFRG, making it often still applicable when other methods fail. For example, complex frustrated and longer-range spin couplings can be treated at no additional numerical cost. Even three dimensional systems pose no particular problem. On the other hand, the PFFRG is limited by the involved approximations which amount to truncating the hierarchy of coupled differential equations for the multi-particle vertex functions. The most common one-loop truncation consists of neglecting the three-particle vertex function and, thus, only allows for the computation of observables that are linear or quadratic in the spin operators. As a result, physical phenomena associated with three-spin or higher correlators are generally not captured. Multiloop approaches which have recently been successfully implemented~\cite{Kugler18_1,kugler18_2,Kiese22,Thoenniss20} are a promising attempt to take into account additional contributions from three-particle vertices.

Among the many advances of recent years, the treatment of finite magnetic fields was previously considered unfeasible due to exceedingly high numerical costs~\cite{Buessen19}.
In this paper, we demonstrate that this generalization is, in fact, viable and numerically not too expensive. Specifically, we develop and apply a PFFRG approach for general spin models of the form
\begin{equation}
\label{eq:GeneralHamiltonian}
    \hat{\mathcal{H}} = \frac{1}{2} \sum_{ij} \sum_{\mu\nu} J^{\mu\nu}_{ij} \hat{S}^{\mu}_{i} \hat{S}^{\nu}_{j} - \sum_{i} \sum_{\mu} h^{\mu}_i \hat{S}^{\mu}_{i},
\end{equation}
with spin-$\frac{1}{2}$ operators $\hat{S}^{\mu}_{i}$ acting on site $i$ and $\mu \in \{x,y,z\}$. The first term describes general anisotropic two-body spin interactions while the second term corresponds to an arbitrary site-dependent Zeeman term $\sim h^{\mu}_{i}$ that explicitly breaks time-reversal symmetry (TRS). We demonstrate the feasibility of this method extension in the absence of any continuous global spin rotation symmetry. However, if a model still exhibits a global U(1) spin rotation symmetry, the complexity of the renormalization group equations is even lower than for a general anisotropic model with TRS.

This method extension enables a variety of new applications. For example, as we show below, the addition of magnetic fields regularizes susceptibility divergences associated with magnetic phase transitions, in analogy to the effects of symmetry breaking charge-density fields or superconducting pairing amplitudes in previous applications of the fermionic FRG~ \cite{Salmhofer04,Gersch05}. The small magnetic seed fields generate finite magnetic order parameters and allows the PFFRG flow to be continued into magnetically ordered phases.
The addition of TRS breaking fields also enables the investigation of field-induced phenomena, such as magnetization plateaus~\cite{Honecker04} or field-induced quantum-paramagnetic phases~\cite{Ivica21,banerjee18}. Finally, since magnetic fields represent one of the simplest external tuning parameter in experiments, our method extension may stimulate further joint theoretical and experimental investigations.

Besides presenting the formal procedure of including magnetic fields in the PFFRG, we also perform some first exploratory applications to reveal the strengths and weaknesses of the method. This is done for some well-studied Heisenberg models on the square, honeycomb and triangular lattices where we can compare PFFRG results with the outcomes of other approaches. Particularly, by regularizing the renormalization group flows of these systems via small magnetic seed fields we obtain their ground state magnetizations at $\Lambda\rightarrow 0$. A recurring observation is that antiferromagnetic magnetizations from PFFRG are larger than the literature values. On the other hand, the precise magnetic ordering patterns are captured correctly even if the seed fields deviate substantially from the expected orders. Furthermore, for the square lattice antiferromagnet we calculate the full magnetization curve up to saturation, which shows good agreement with other methods~\cite{Honecker04}. We discuss in detail possible reasons for these strengths and weaknesses.

The paper is structured as follows.
A brief outline of the key features of the PFFRG method in its previous formulation is presented in Sec.~\ref{sec:PFFRG}, as preparation for the following, more formal sections. In Sec.~\ref{sec:Symmetries}, we discuss the symmetry properties of the fermionic self-energy and two-particle vertex functions. An efficient, symmetry-constrained vertex parameterization is presented in Sec.~\ref{sec:Parametrization}, which is then applied in Sec.~\ref{sec:FlowEquation} to obtain the PFFRG flow equations for arbitrary spin models with linear and bilinear spin terms. In Sec.~\ref{sec:Application}, we apply our generalized PFFRG approach to square, honeycomb and triangular lattice Heisenberg models and investigate their magnetically ordered phases at $T=0$.
We conclude the paper with a discussion of the method in Sec.~\ref{sec:Discussion}.

\section{\label{sec:PFFRG} PFFRG in a nutshell}

In this section, we sketch some basic properties of the PFFRG, before we formulate the method for TRS-breaking systems. 
For a more extensive introduction of the PFFRG we refer the reader to Ref.~\cite{Mueller23}.
 
Within PFFRG, $S=\frac{1}{2}$ spin operators are mapped onto pseudo-fermion operators via~\cite{Abrikosov65,Affleck88}
\begin{equation}
\label{eq:pseudoFermionMapping}
    \hat{S}_{i}^{\mu} \rightarrow \frac{1}{2} \sum_{\alpha, \beta = \uparrow, \downarrow} \sigma^{\mu}_{\alpha\beta} \hat{f}^{\dagger}_{i\alpha} \hat{f}_{i\beta},
\end{equation}
where $\sigma^{\mu}_{\alpha\beta}$ are Pauli matrix entries and $\hat{f}^{\dagger}_{i\alpha}$ ($\hat{f}_{i\alpha}$) is a creation (annihilation) operator of a fermion on site $i$ with spin $\alpha\in\{\uparrow,\downarrow\}$. Employing this replacement in the Hamiltonian $\hat{\mathcal{H}}$ of Eq.~(\ref{eq:GeneralHamiltonian}) results in the pseudo-fermionic Hamiltonian
\begin{align}
    \hat{\mathcal{H}}_{\text{pf}} = &\frac{1}{8} \sum_{ij} \sum_{\mu \nu} J_{ij}^{\mu \nu} \sum_{\substack{\alpha \beta\\ \alpha' \beta'}} 
     \sigma^{\mu}_{\alpha \beta} \sigma^{\nu}_{\alpha' \beta'}
    \hat{f}_{i \alpha}^{\dagger} \hat{f}_{i \beta} \hat{f}_{j \alpha'}^{\dagger} \hat{f}_{j \beta'} \nonumber\\
    &-\frac{1}{2}\sum_{i} \sum_{\mu} h_{i}^{\mu}   \sum_{\alpha\beta}\sigma^{\mu}_{\alpha\beta} \hat{f}_{i \alpha}^{\dagger} \hat{f}_{i \beta}.
    \label{eq:PseudoFermionHamiltonian}
\end{align}
As a variant of the FRG, the PFFRG yields coupled differential equations for the $\Lambda$-dependent fermionic $n$-particle vertex functions called flow equations. Specifically, after introducing the renormalization group parameter $\Lambda$ in the free single-particle Green function, the flow equations are derived by taking the $\Lambda$-derivative of the generating functional for two-particle-irreducible vertex functions~\cite{Metzner12}. The resulting infinite set of coupled flow equations is formally exact. However, as the flow equation for the $n$-particle vertex contains the $n+1$-particle vertex, the infinite set of coupled differential equations has to be truncated to obtain a finite set that can be solved numerically. In the most common variant of the PFFRG, this truncation is performed on the so called one-loop plus Katanin level, where three-particle and higher vertices are set to zero except for certain vertex contributions (the Katanin terms) that yield a self-consistent feedback of the self-energy into the two-particle vertex. In total, this results in coupled flow equations for the self-energy $\Sigma^{\Lambda}(1'|1)$ and the two-particle vertex $\Gamma^{\Lambda}(1',2'|1,2)$. In a diagrammatic representation, $1$ and $2$ ($1'$ and $2'$) are the arguments of the incoming (outgoing) fermion lines, consisting of Matsubara frequency $\omega$, spin $\alpha$ and site argument $i$, i.e. $1=\{\omega_1,\alpha_1,i_1\}$. The explicit equations are given by
\begin{equation}
\label{eq:BasicFlowEq1}
    \frac{d}{d\Lambda} \Sigma^{\Lambda}(1'|1) = - \frac{1}{2\pi} \sum_{2',2} \Gamma^{\Lambda}(1',2'|1,2) \mathcal{S}^{\Lambda}(2|2')
\end{equation}
and
\begin{widetext}
\begin{align}    
\frac{d}{d\Lambda} \Gamma^{\Lambda}(1',2'|1,2) =& \frac{1}{2 \pi} \sum_{\substack{3',4',\\3,4}} \Big[
\Gamma^{\Lambda}(1',2'|3,4)\Gamma^{\Lambda}(3',4'|1,2) - \Gamma^{\Lambda}(1',4'|1,3)\Gamma^{\Lambda}(3',2'|4,2) -  \Gamma^{\Lambda}(1',3'|1,4)\Gamma^{\Lambda}(4',2'|3,2)\nonumber\\
&+ \Gamma^{\Lambda}(2',4'|1,3)\Gamma^{\Lambda}(3',1'|4,2) +  \Gamma^{\Lambda}(2',3'|1,4)\Gamma^{\Lambda}(4',1'|3,2) \Big] \cdot G^{\Lambda}(3|3')\tilde{\mathcal{S}}^{\Lambda}(4|4'),\label{eq:BasicFlowEq2}
\end{align}
\end{widetext}
where $G^{\Lambda}(1'|1)$ is the full single-particle Green function, $\mathcal{S}^{\Lambda}(1|1') = -\frac{d}{d \Lambda}G^{\Lambda}(1|1')|_{\Sigma^\Lambda=\rm{const}}$ is the single-scale propagator (where the $\Lambda$-derivative ignores the $\Lambda$-dependence of $\Sigma^\Lambda$) and $\tilde{\mathcal{S}}^{\Lambda}(1|1') = -\frac{d}{d \Lambda}G^{\Lambda}(1|1')$ is the single-scale propagator in Katanin approximation.
So-called higher-loop truncations have also been recently applied which generate additional vertex contributions to $\Sigma^{\Lambda}(1'|1)$ and $\Gamma^{\Lambda}(1',2'|1,2)$ not contained in Eqs.~(\ref{eq:BasicFlowEq1}) and (\ref{eq:BasicFlowEq2}) \cite{Kiese22,Thoenniss20}.

In PFFRG, the flow parameter $\Lambda$ is introduced as an infrared frequency cutoff that suppresses the low-frequency part $|\omega| \lesssim\Lambda$ of the free single-particle propagator $G_0$. In this paper, we apply a sharp frequency cutoff, realized by the replacement $G_0\rightarrow \theta(|\omega|-\Lambda)G_0\equiv G_0^\Lambda$. Recent works have also used smooth cutoff functions to avoid possible artefacts from the non-analyticity of the step function~\cite{Thoenniss20, Kiese22,Ritter22}. The flow equations are solved numerically from the known infinite cutoff limit $\Lambda \rightarrow \infty$, where only the bare parameters of the Hamiltonian enter, towards the physical cutoff-free limit $\Lambda=0$, e.g. by applying a Runge-Kutta method. Intuitively, the PFFRG cutoff parameter shares some properties with the temperature. Like in a cooling process from high to low temperatures, the renormalization group flow starts in the paramagnetic phase at large $\Lambda$ and may (or may not) sense a magnetic phase transition as $\Lambda$ is lowered.

The pseudo-fermion mapping in Eq.~(\ref{eq:pseudoFermionMapping}), inherent to the method, doubles the Hilbert space dimension of each lattice site by introducing unphysical states. A crucial challenge in PFFRG is to avoid spurious contributions of these unphysical states in the numerical results. The physical spin-$\frac{1}{2}$ states of a lattice site are characterized by single fermion occupancy. If no further measures are taken to enforce this single-particle constraint in Eqs.~(\ref{eq:BasicFlowEq1}) and (\ref{eq:BasicFlowEq2}), it is still fulfilled {\it on average}. This approach, which we also pursue here, has been successfully applied at zero temperature. However, at finite temperatures, the impact of unphysical states rapidly increases~\cite{Schneider23}, making the method inapplicable. A better fulfillment of the single-particle constraint at finite temperatures can be traded against increased numerical costs by applying the Popov-Fedotov trick~\cite{Schneider23}. Alternatively, additional terms in the Hamiltonian can be introduced which energetically penalize the unphysical states~\cite{Baez17,HagymasiS1PC22}. The recently developed pseudo-Majorana FRG and spin-FRG circumvent the problem of unphysical states by either applying a Majorana fermion mapping absent of unphysical states~\cite{Niggemann21,Niggemann22}, or by applying the FRG formalism directly to a spin model~\cite{Krieg19, Goll19}.

The physical observables calculated with the PFFRG are spin-spin correlators and spin susceptibilities obtained from $\Sigma^{\Lambda}(1'|1)$ and $\Gamma^{\Lambda}(1',2'|1,2)$. These magnetic response functions are computed from high cutoff scales $\Lambda\rightarrow\infty$ down to the cutoff-free limit $\Lambda=0$. Since, by construction, Eqs.~(\ref{eq:BasicFlowEq1}) and (\ref{eq:BasicFlowEq2}) respect all symmetries of the considered Hamiltonian, a magnetic phase transition associated with spontaneously broken symmetries enforces a flow breakdown at a critical cutoff $\Lambda_{\rm{c}}$. Ideally, this breakdown would be signalled by a divergence of the susceptibility at the respective ordering wave vector. However, further applied approximations concerning the maximum considered correlation distance and the frequency resolution of vertex functions are observed to suppress the divergence, which instead appears as a kink in the susceptibility flow (see Sec.~\ref{sec:Application}). While the momentum dependent susceptibility just above the kink allows to identify the type of magnetic order, results obtained at cutoff scales $\Lambda$ below this breakdown feature are unphysical. On the other hand, a susceptibility flow down to $\Lambda=0$ without any kink-like features signifies a non-magnetic ground state.

\section{\label{sec:Symmetries}Symmetry properties of Green and vertex functions}

\begin{table*}[t]
\renewcommand{\arraystretch}{2}
\begin{tabular}{|c|c|c|} 
 \hline
 Symmetry  & $G(1'|1)$ & $G(1',2'|1,2)$  \\ 
 \hline\hline
 Particle exchange & Not applicable & $G(1',2'|1,2)=-G(1',2'|2,1)=-G(2',1'|1,2)$ \\
 \hline 
 $W = i\sigma^{z}$ & $G(1'|1)\sim \delta_{i_{1'}i_{1}}$& 
 $G(1',2'|1,2) = G_{||}(1',2'|1,2) \delta_{i_{1'}i_{1}} \delta_{i_{2'}i_{2}}$\\
 (gauge transformation)&&
        $\phantom{G(1',2'|1,2)}+ G_{\times}(1',2'|1,2) \delta_{i_{2'}i_{1}} \delta_{i_{1'}i_{2}}$
 \\
 \hline
 $W = i\sigma^{x}$ or $W = i\sigma^{y}$& $G(1'|1)=-\alpha_{1}\alpha_{1'} G(-\bar{1}|-\bar{1}')$ &$G_{||}(1',2'|1,2)=-\alpha_{1}\alpha_{1'} G_{||}(-\bar{1},2'|-\bar{1'},2)$\\
 (gauge transformation)&&$\phantom{G_{||}(1',2'|1,2)}=-\alpha_{2}\alpha_{2'} G_{||}(1',-\bar{2}|1,-\bar{2'})$\\\hline
 Hermiticity & $G(1'|1)=G(-1|-1')^{*}$& $G(1',2'|1,2)=G(-1,-2|-1',-2')^{*}$  \\
 \hline
 Time-translational invariance & $G(1'|1) \sim\delta(\omega_{1'}-\omega_{1})$& $G(1',2'|1,2)\sim \delta(\omega_{1'}+\omega_{2'}-\omega_{1}-\omega_{2})$   \\
 \hline
\end{tabular}
\caption{Symmetry properties of the pseudo-fermion single-particle Green function $G(1'|1)$ (second column) and two-particle Green function $G(1',2'|1,2)$ (third column) enforced by the symmetries listed in the first column. Number arguments denote a set of variables, containing the Matsubara frequency $\omega$, spin $\alpha$ and site $i$, i.e., $1=\{\omega_{1}, \alpha_{1}, i_{1}\}$. Spin variables take values $\alpha=\pm 1$ which is used interchangeably with $\alpha=\uparrow,\downarrow$. Furthermore, we define $-1=\{-\omega_{1}, \alpha_{1}, i_{1}\}$ and $\bar{1}=\{\omega_{1}, -\alpha_{1}, i_{1}\}$. The corresponding symmetry relations for the self-energy $\Sigma$ and the two-particle vertex $\Gamma$ are obtained by simply replacing $G\leftrightarrow\Sigma$ and $G\leftrightarrow \Gamma$ in the second and third columns, respectively.}
\label{tab:Symmetries}
\end{table*}

Vertex functions obtained as solutions of the flow equations (and the related Green functions) respect all symmetries of the Hamiltonian exactly throughout the $\Lambda$ flow. While flow equations could in principle be solved directly from Eqs. \eqref{eq:BasicFlowEq1} and \eqref{eq:BasicFlowEq2} by evaluating the right-hand side of the equations for any vertex argument combination of the left-hand side, such an approach is numerically inefficient. For an efficient numerical solution it is crucial to take into account a model's symmetries and build them directly into the vertex parameterizations to avoid redundant computations.

A detailed derivation of the symmetry properties of Green and vertex functions for the time-reversal symmetric PFFRG has already been performed in Ref.~\cite{Buessen19}. In the following, we repeat this approach for the case where no time-reversal symmetry is imposed. The resulting symmetry constraints for Green and vertex functions are collected in Table~\ref{tab:Symmetries} and form the basis for our spin parameterizations discussed in Sec.~\ref{sec:Parametrization}.

\subsection{Action of symmetry operations on Green and vertex functions}

Symmetries of a pseudo-fermion Hamiltonian $\hat{\mathcal{H}}_{\rm{pf}}$ carry over to the fermionic vertex functions non-trivially. Consider a general symmetry of $\hat{\mathcal{H}}_{\rm{pf}}$ described by the symmetry operator $\hat{w}$; then the commutation relation
\begin{align}
    [\hat{\mathcal{H}}_{\rm{pf}},\hat{w}] = 0
\end{align}
must hold. Applying the symmetry operation $\hat{w}$ to the imaginary time-ordered $n$-particle Green function defined by \cite{NegeleOrland}
\begin{equation}
\begin{split}
        &G(1',2',\ldots,n'|1,2,\ldots,n) = -\hspace*{-2pt}\int_0^{\infty}\hspace*{-6pt}d\tau_{1'}\cdots d\tau_{n'}d\tau_{1}\cdots d\tau_{n}\times\\ 
        &\exp[i(\tau_{1'}\omega_{1'}+\ldots+\tau_{n'}\omega_{n'}-\tau_{1}\omega_{1}-\ldots-\tau_{n}\omega_{n})]\times\\
        &\langle \mathcal{T}_{\tau} \big( \hat{f}_{i_{1'}\alpha_{1'}}(\tau_{1'})\cdots\hat{f}_{i_{n'}\alpha_{n'}}(\tau_{n'}) \hat{f}^{\dagger}_{i_{n}\alpha_{n}}(\tau_{n})\cdots \hat{f}^{\dagger}_{i_{1}\alpha_{1}}(\tau_{1}) \big ) \rangle\label{eq:green_general}
\end{split}
\end{equation}
corresponds to transforming each fermion operator according to
\begin{equation}
\hat{f}^{(\dagger)}_{i\alpha}(\tau)\rightarrow \hat{w}\hat{f}^{(\dagger)}_{i\alpha}(\tau)\hat{w}^\dagger,
\end{equation}
where $\mathcal{T}_{\tau}$ is the time-ordering operator. Note that at finite temperatures, the upper integral limit in Eq.~(\ref{eq:green_general}) is $1/T$, which in our present zero-temperature considerations becomes infinite. Requiring that the $n$-particle Green functions remain unchanged under this symmetry operation leads to constrained functional dependencies in the variables $1',\ldots,n'$ and $1,\ldots,n$, which we discuss below.

Next, one needs to translate the symmetry constraints for the Green functions into symmetry constraints for the vertex functions $\Sigma(1'|1)$ and $\Gamma^{\Lambda}(1',2'|1,2)$ since PFFRG is formulated in terms of these latter objects. This is accomplished by the known relations between both, such as Dyson's equation
\begin{equation}
\label{eq:DysonEq}
    G(1'|1) = \left(G_0^{-1}- \Sigma\right)^{-1}(1'|1).
\end{equation}
Here, the inversions $(\cdots)^{-1}$ denote matrix inversions in frequency, spin, and site variables. Furthermore, the fermionic two-particle vertex $\Gamma(1',2'|1,2)$ is related to single and two-particle Green functions $G(1'|1)$ and $\Gamma(1',2'|1,2)$ via the tree expansion~\cite{NegeleOrland}
\begin{align}
    &G(1',2'|1,2) = \nonumber\\
    &- \sum_{3',4',3,4} G(1'|3') G(2'|4') G(3|1) G(4|2) \Gamma(3',4'|3,4)\nonumber\\ &+ G(1'|1)G(2'|2) - G(2'|1)G(1'|2).\label{eq:TreeDiagram}
\end{align}
In our case, one finds that all symmetries transform $G(1'|1)$ and $\Sigma(1'|1)$ equivalently and the same applies to $G(1',2'|1,2)$ and $\Gamma(1',2'|1,2)$.

In the following, we discuss the individual symmetries of $\hat{\mathcal{H}}_{\rm{pf}}$ and the resulting restrictions on Green functions more explicitly, making use of the properties outlined here. The corresponding vertex function symmetries follow trivially. One can generally distinguish between two types of symmetries, those of the original spin Hamiltonian $\hat{\mathcal{H}}$ and those which only arise from the fermion mapping. The latter one corresponds to a local SU(2) gauge symmetry of the pseudo-fermion Hamiltonian $\hat{\mathcal{H}}_{\rm{pf}}$ that will be discussed first, see Sec.~\ref{sec:gauge_symmetries}. Additionally, as specified in Table~\ref{tab:Symmetries}, the fermion statistics leads to a minus sign whenever two particles are exchanged ($1\leftrightarrow 2$ or $1'\leftrightarrow 2'$) and Hermiticity of $\hat{\mathcal{H}}_{\rm{pf}}$ poses restrictions under complex conjugation. 

\subsection{Gauge symmetries of $\hat{\mathcal{H}}_{\rm{pf}}$}\label{sec:gauge_symmetries}
The pseudo-fermion mapping in Eq.~\eqref{eq:pseudoFermionMapping} can be rewritten as \cite{Affleck88}
\begin{equation}
    \hat{S}^{\mu}_{i} \rightarrow \frac{1}{4} \text{tr}(\hat{F}^{\dagger}_{i}\sigma^{\mu}\hat{F}_{i}),
\end{equation}
with
\begin{equation}
\hat{F}_{i} =
    \begin{pmatrix}
\hat{f}_{i\uparrow} & \hat{f}^{\dagger}_{i\downarrow} \\
\hat{f}_{i\downarrow} & -\hat{f}^{\dagger}_{i\uparrow}
\end{pmatrix}.
\end{equation}
This expression makes it apparent that the pseudo-fermionic mapping introduces a local SU(2) gauge symmetry $\hat{w}_i$ defined by the operation
\begin{equation}
    \hat{F}_{i} \rightarrow  \hat{w}_i\hat{F}_{i}\hat{w}_i^\dagger\equiv \hat{F}_{i}{ W}_i,\label{eq:gauge_trafo}
\end{equation}
under which $\hat{S}_i^\mu$ remains invariant. Here, $W_i$ is an arbitrary site-dependent unitary $2\times 2$ matrix. In the following, we discuss the constraints that arise in the special cases when a gauge transformation $W_i$ is only applied on a single site $j$, where we implement it as $W_{j}= i\sigma^{\mu}$ with $\mu=x,y,z$. No transformation is applied to the other sites.

We first consider a gauge transformation $W_{j} = i\sigma^{z}$ under which fermion operators of site $j$ transform as
\begin{equation}
    \begin{pmatrix}
    \hat{f}_{j\alpha} \\
    \hat{f}^{\dagger}_{j\alpha}
    \end{pmatrix} \longrightarrow 
        i \begin{pmatrix}
    \hat{f}_{j\alpha} \\
    -\hat{f}^{\dagger}_{j\alpha}
    \end{pmatrix}.
\end{equation}
Requiring that Green functions stay invariant under this operation, the phase factor $i$ from transforming $\hat{f}_{j\alpha}$ has to be cancelled by the phase $-i$ from transforming $\hat{f}_{j\alpha}^\dagger$ {\it on the same site} $j$. This is only possible if $n$-particle Green function site arguments of incoming fermion lines $\{i_{1},\cdots,i_{n}\}$ are given by a permutation of the site arguments of outgoing fermion lines $\{i_{1'},\cdots,i_{n'}\}$.
For the single-particle Green function this simply means $G(1'|1)\sim \delta_{i_{1'}i_1}$ while for the two-particle Green function, this enforces the parameterization
\begin{align}
        G(1',2'|1,2) = &G_{||}(1',2'|1,2) \delta_{i_{1'}i_{1}} \delta_{i_{2'}i_{2}} \nonumber\\
        +& G_{\times}(1',2'|1,2) \delta_{i_{2'}i_{1}} \delta_{i_{1'}i_{2}}.\label{eq:2GPara}
\end{align}
Thus, $G(1',2'|1,2)$ can only depend on two site arguments $i_1$ and $i_2$. The application of particle exchange further reveals the relation
\begin{equation}
\label{eq:2GParaRelation}
    G_{\times}(1',2'|1,2) = -G_{||}(1',2'|2,1).
\end{equation}

Next, the gauge transformations $W_{j} =i\sigma^{x}$ and $W_{j} =i\sigma^{y}$ transform the fermion operators as
\begin{equation}
    \begin{pmatrix}
    \hat{f}_{j\alpha} \\
    \hat{f}^{\dagger}_{j\alpha}
    \end{pmatrix} \longrightarrow 
    i \begin{pmatrix}
    \alpha\hat{f}^{\dagger}_{j\bar{\alpha}} \\
    \bar{\alpha}\hat{f}_{j\bar{\alpha}}
    \end{pmatrix}
\end{equation}
and 
\begin{equation}
    \begin{pmatrix}
    \hat{f}_{j\alpha} \\
    \hat{f}^{\dagger}_{j\alpha}
    \end{pmatrix} \longrightarrow 
    -\alpha \begin{pmatrix}
    \hat{f}^{\dagger}_{j\bar{\alpha}} \\
    \hat{f}_{j\bar{\alpha}}
    \end{pmatrix},
\end{equation}
respectively. Here, spin variables are understood as $\alpha=\pm 1$ (where $\bar{\alpha}=-\alpha$) which is used interchangeably with $\alpha=\uparrow,\downarrow$. Both transformations exchange the roles of creation and annihilation operators. The implied restrictions for single and two-particle Green functions are equivalent for both transformations and are given in Table~\ref{tab:Symmetries}. These restrictions are formulated in terms of $G_{||}(1',2'|1,2)$, for which it is known that $i_{1'}=i_{1}$ and $i_{2'}=i_{2}$.

Note that, since the gauge transformations are local, they can be applied to either of the two sites $i_1$ and $i_2$ involved in $G_{||}(1',2'|1,2)$.

\subsection{\label{sec:PhysicalSymmetries}Physical symmetries of $\hat{\mathcal{H}}$}

Since we only consider spin systems in equilibrium, time-translation symmetry holds. As specified in Table~\ref{tab:Symmetries}, this leads to energy conservation which implies that the sum of frequencies on incoming fermion lines equals the sum of frequencies on outgoing fermion lines.

In the absence of any global spin rotation symmetries, the only remaining symmetries which the general spin Hamiltonian $\hat{\mathcal{H}}$ in Eq.~(\ref{eq:GeneralHamiltonian}) may have are lattice symmetries. Their consequences are not listed in Table~\ref{tab:Symmetries}, but are explained here in words. 
Lattice symmetries enforce relations between Green functions with different site arguments. In the simplest (and also most common case), a lattice model allows each site to be mapped onto every other site by a combination of translation, point group and global spin rotation symmetry operations, i.e., all sites are symmetry equivalent. In that case, knowing $G(1'|1)$ on one arbitrary reference site $i_1$ specifies it for all other sites. Similarly, one site index of $G_{||}(1',2|1,2)$, e.g., $i_1$, can always be set to the reference site. After fixing $i_1$, the set of sites $i_2$ for which $G_{||}(1',2|1,2)$ constitute independent functions is further reduced by point group symmetries. If a spin system has many symmetry-inequivalent lattice sites, the Green functions $G(1'|1)$ and $G_{||}(1',2|1,2)$ are independent for each of these reference sites $i_1$.

For spin-anisotropic Hamilonians, the symmetry operations may consist of combined lattice \emph{and} spin transformations. If such a combined lattice and spin symmetry occurs in the presence of (site-dependent) magnetic fields, one needs to take into account that spin rotations act explicitly on the single-particle Green function. This property will be discussed in more detail in the next section.

\section{\label{sec:Parametrization}Parameterization of spin dependence of vertex functions}

We have now collected all symmetry constraints and can propose an efficient parameterization of the spin structure of Green and vertex functions. Specifically, we will follow previous approaches \cite{Reuther10,Hering17,Buessen19} and express the spin dependencies in terms of Pauli and identity matrices. This leads to a formulation where the self-energy $\Sigma$ (the two-particle vertex $\Gamma$) is parameterized by a linear (bilinear) combination of these matrices. The parameterization of the spin structure of Green and vertex functions in previous time-reversal symmetric PFFRG implementations will be extended to capture components that become finite in the absence of TRS. As will be shown, only the parameterizations for the single-particle Green function and self-energy have to be adjusted to incorporate the broken TRS. The convenient property of purely real or purely imaginary vertex components will remain such that the numerical effort for solving the renormalization group equations stays moderate.

\subsection{Self-energy parameterization}

The presence of a local gauge freedom with $W_i=i \sigma^{z}$ [see Eq.~(\ref{eq:gauge_trafo})] in combination with time-translational invariance allows us to write the self-energy as
\begin{equation}
\Sigma(1'|1) = \Sigma_{i_1}(\omega_{1},\alpha_{1'},\alpha_{1}) \delta_{i_{1'}i_{1}} \delta(\omega_{1'}-\omega_{1}).\label{eq:GreenFctLocalPara}
\end{equation}
The spin dependence of the remaining function $\Sigma_{i}(\omega_{1},\alpha_{1'},\alpha_{1})$ can be expanded as
\begin{align}
&\Sigma_{i}(\omega_{1},\alpha_{1'},\alpha_{1})=\sum_{\rho} \Sigma^{\rho}_{i}(\omega_{1}) \sigma^{\rho}_{\alpha_{1'}\alpha_{1}}\label{eq:PropagatorPara2}\\
&=-i \gamma^{0}_{i}(\omega_{1}) \delta_{\alpha_{1'}\alpha_{1}} +  \sum_{\mu} \gamma^{\mu}_{i}(\omega_{1}) \sigma^{\mu}_{\alpha_{1'}\alpha_{1}}\label{eq:PropagatorPara3},
\end{align}
with $\sigma^{\rho}$ being either a Pauli matrix for $\rho \in \{x,y,z \}$ or the $2\times 2$ identity matrix for $\rho=0$. Here and in the following, we use the convention that indices $\mu$, $\nu$ capture the Cartesian coordinates $\mu,\nu\in\{x,y,z\}$ while $\rho$ and $\varphi$ also contain the zeroth component, $\rho,\varphi\in\{0,x,y,z\}$.
The gauge freedom from $W_i=i \sigma^{x}$ leads to even or odd frequency structures of the self-energy components
\begin{equation}
\begin{split}
    \gamma^{0}_{i}(\omega) =& -\gamma^{0}_{i}(-\omega), \\
    \label{eq:PropagatorSymmetry1}
    \gamma^{\mu}_{i}(\omega) =& \gamma^{\mu}_{i}(-\omega).
    \end{split}
\end{equation}
Finally, Hermiticity enforces $\gamma^{\rho}(\omega)$ to be purely real
\begin{equation}
\begin{split}
    \label{eq:PropagatorSymmetry2}
    \gamma^{\rho}_{i}(\omega) &\in \mathbb{R}.
    \end{split}
\end{equation}
An analogous parameterization can be applied to the single-particle Green function $G(1'|1)$ as follows
\begin{align}
&G(1'|1)=\sum_{\rho} G^{\rho}_{i_1}(\omega_{1}) \sigma^{\rho}_{\alpha_{1'}\alpha_{1}}\delta_{i_{1'}i_{1}} \delta(\omega_{1'}-\omega_{1})\nonumber\\
&=\Big(-i g^{0}_{i_1}(\omega_{1}) \delta_{\alpha_{1'}\alpha_{1}} +  \sum_{\mu} g^{\mu}_{i_1}(\omega_{1}) \sigma^{\mu}_{\alpha_{1'}\alpha_{1}}\Big)\times\nonumber\\
&\delta_{i_{1'}i_{1}} \delta(\omega_{1'}-\omega_{1}).\label{eq:green_parameterization}
\end{align}
Equivalent symmetries for $g_i^{\rho}(\omega)$ and $\gamma_i^{\rho}(\omega)$ apply. 
Formulating Dyson's equation [see Eq.~\eqref{eq:DysonEq}] in terms of $g_i^{\rho}(\omega)$ and $\gamma_i^{\rho}(\omega)$ leads to
\begin{align}
\begin{split}
\label{eq:PropSERelation}
    &g_{i}^{0}(\omega) = \frac{\omega+\gamma_{i}^{0}(\omega)}{(\omega+\gamma_{i}^{0}(\omega))^{2}+\sum_{\nu}(\gamma_{i}^{\nu}(\omega))^{2}}, \\
    &g_{i}^{\mu}(\omega) = \frac{-\gamma_{i}^{\mu}(\omega)}{(\omega+\gamma_{i}^{0}(\omega))^{2}+\sum_{\nu}(\gamma_{i}^{\nu}(\omega))^{2}}.
    \end{split}
\end{align}
The different components of $g_i^{\rho}$ and $\gamma_i^{\rho}$ have distinct transformation properties under spin rotations and time reversal~\cite{Buessen19}. The zeroth components $g_i^0$ and $\gamma_i^0$ behave like a scalar, thus, they remain invariant under spin rotations. On the other hand, the Cartesian components $\mu\in\{x,y,z\}$ transform as the components of a pseudovector. Specifically, for a spin rotation $\hat{S}_i^\mu\rightarrow \sum_\nu R^{\mu\nu}\hat{S}_i^\nu$ (with an orthogonal $3\times 3$ matrix $R$) they transform as $g_i^\mu\rightarrow \sum_\nu R^{\mu\nu}g_i^\nu$ and $\gamma_i^\mu\rightarrow \sum_\nu R^{\mu\nu}\gamma_i^\nu$. Furthermore, under time reversal and the frequency property in Eq.~(\ref{eq:PropagatorSymmetry1}) all Cartesian components change sign, $g_i^\mu\rightarrow -g_i^\mu$ and $\gamma_i^\mu\rightarrow -\gamma_i^\mu$, in analogy to the behavior of spin operators $\hat{S}_i^\mu$ under time-reversal. From the transformation behavior under time-reversal and Eq.~(\ref{eq:PropagatorSymmetry1}) it also follows that $g_i^\mu=\gamma_i^\mu=0$ for systems with TRS.

The transformation properties of $\gamma_{i}^{\mu}$ (and $g_{i}^{\mu}$) need to be taken into account when performing symmetry operations. Particularly, if a TRS-broken system is invariant under a combined lattice and spin rotation symmetry, $\gamma_{i}^{\mu}$ and $\gamma_{j}^{\mu}$ can be different on two lattice sites $i\neq j$, even though $i$ and $j$ are symmetry equivalent. On the other hand, the zeroth component of the self-energy is always equal on symmetry-equivalent sites $i$ and $j$, i.e., $\gamma_i^0=\gamma_j^0$. 

\subsection{Two-particle vertex parameterization}

The parameterization of the two-particle vertex is carried out similarly. First, Eqs.~(\ref{eq:2GPara}) and (\ref{eq:2GParaRelation}) in combination with time-translation invariance imply the parameterization
\begin{equation}
\label{eq:gammaparallpara}
\begin{split}
    &\Gamma(1',2'|1,2) = (\Gamma_{||}(1',2'|1,2) \delta_{i_{1'}i_{1}}\delta_{i_{2'}i_{2}} \\&-\Gamma_{||}(1',2'|2,1)\delta_{i_{1'}i_{2}}\delta_{i_{2'}i_{1}}) \delta(\omega_{1}+\omega_{2}-\omega_{1'}-\omega_{2'}).
    \end{split}
\end{equation}
Next, the dependence of $\Gamma_{||}(1',2'|1,2)$ on spin indices is written in the most general form as a bilinear expansion in Pauli and identity matrices
\begin{equation}
\label{eq:TPVertexSigmaPara}
    \Gamma_{||}(1',2'|1,2) = \sum_{\rho,\varphi}  \Gamma^{\rho\varphi}_{i_1 i_2}(\omega_{1'},\omega_{2'}|\omega_{1},\omega_{2}) \sigma_{\alpha_{1'}\alpha_{1}}^{\rho} \sigma_{\alpha_{2'}\alpha_{2}}^{\varphi}.
\end{equation}
The functions $\Gamma^{\rho\varphi}_{i_1 i_2}(\omega_{1'},\omega_{2'}|\omega_{1},\omega_{2})$ fulfill various useful relations. Specifically, from the gauge freedom $W_i=i \sigma^{x}$ or $W_i=i \sigma^{y}$, combined with Hermiticity, one finds
\begin{equation}
    \Gamma^{\rho\varphi} \in \begin{cases}
       \mathbb{R} &  \text{if}\ \rho,\varphi=0  \ \text{or}\ \rho,\varphi \neq 0 \ \\
        i\mathbb{R} & \text{otherwise}.
    \end{cases}
\end{equation}
The remaining symmetries of Table \ref{tab:Symmetries} manifest themselves in the properties
\begin{align}
    \Gamma^{\rho\varphi}_{i_{1}i_{2}}(s,t,u) =& \Gamma^{\rho\varphi}_{i_{1}i_{2}}(-s,t,-u)^{*} \label{eq:tu-symmetry}\\
    =& \Gamma^{\varphi\rho}_{i_{2}i_{1}}(s,-t,-u) \\
    =& (-1)^{\delta_{\varphi0}} \Gamma^{\rho\varphi}_{i_{1}i_{2}}(u,t,s).
\end{align}
Here, to incorporate energy conservation, we have switched to three frequency arguments, using the definitions 
\begin{equation}
        s = \omega_{1'} + \omega_{2'},\quad
        t =  \omega_{1'} - \omega_{1},\quad
        u =  \omega_{1'} - \omega_{2}.
\end{equation}
It is worth emphasizing that {\it all} components $\rho,\varphi\in\{0,x,y,z\}$ of $\Gamma^{\rho\varphi}$ can already be finite when TRS is preserved~\cite{Buessen19}. In other words, the generalization to TRS-broken systems does not generate any new components of $\Gamma^{\rho\varphi}$. However, the breaking of TRS removes the property
\begin{equation}
\label{eq:TRSFrequencySymmetry}
    \Gamma^{\rho\varphi}_{i_{1}i_{2}}(s,t,u) = \Gamma^{\rho\varphi}_{i_{1}i_{2}}(-s,-t,-u)
\end{equation}
which holds when TRS is intact~\cite{Buessen19}. The absence of this frequency symmetry doubles the number of frequency arguments for which the two-particle vertex function has to be evaluated when solving the flow equations. However, as will be discussed in Sec.~\ref{sec:FlowEquation}, the main reason for increased numerical complexity when TRS is broken are the additional terms $\gamma_i^\mu\neq 0$ in the parameterization of the self-energy.

\section{\label{sec:FlowEquation}Flow equations for TRS-broken systems}
\subsection{Explicit flow equations}
We have now finished all the preparatory work to present the PFFRG flow equations for $\Sigma_{i}^{\rho,\Lambda}(\omega)$ [see Eq.~(\ref{eq:PropagatorPara2})] and $\Gamma_{i_1i_2}^{\rho\varphi,\Lambda}(s,t,u)$ [see Eq.~(\ref{eq:TPVertexSigmaPara})] for the general Hamiltonian in Eq.~(\ref{eq:GeneralHamiltonian}). These equations are obtained by inserting the parameterizations from Eqs.~\eqref{eq:GreenFctLocalPara}, \eqref{eq:PropagatorPara2}, \eqref{eq:gammaparallpara}, and \eqref{eq:TPVertexSigmaPara} into Eqs. \eqref{eq:BasicFlowEq1} and \eqref{eq:BasicFlowEq2}. They read as
\begin{align}
&\frac{d}{d\Lambda} \Sigma^{\rho,\Lambda}_{i}(\omega) =\nonumber\\
&\frac{1}{4\pi} \int d\omega' \Big[-4 \sum_{j} \sum_{a} \Gamma^{\rho a,\Lambda}_{i j}(\omega+\omega',0,\omega-\omega') \mathcal{S}^{a,\Lambda}_{j}(\omega')\nonumber \\
&+ \sum_{abc} \Gamma^{ab,\Lambda}_{ii}(\omega+\omega',\omega-\omega',0)  \mathcal{S}^{c,\Lambda}_{i}(\omega') \text{tr}(\sigma^{a}\sigma^{c}\sigma^{b}\sigma^{\rho}) \Big]\label{eq:SEFlowequationImplicit}
\end{align}
and
\begin{widetext}
\begin{align}
\label{eq:FlowEqComplete2}
    \begin{split}
        &\frac{d}{d\Lambda}\Gamma^{\rho\varphi,\Lambda}_{i_1 i_2}(s,t,u) = \frac{1}{8\pi} \int d\omega' \sum_{abcdef}\\
        \Big[ &\Gamma^{ab,\Lambda}_{i_1 i_2}(s,-\omega'-\omega_{2'},\omega_{1'}+\omega') \Gamma^{cd,\Lambda}_{i_1 i_2}(s,\omega_{2}+\omega',\omega_{1}+\omega')
         \big(G^{e,\Lambda}_{i_{1}}(s+\omega')\tilde{\mathcal{S}}^{f,\Lambda}_{i_{2}}(\omega')^{*}+G^{f,\Lambda}_{i_{2}}(\omega')^{*} \tilde{\mathcal{S}}^{e,\Lambda}_{i_{1}}(s+\omega') \big)\times\\
        &\text{tr}\big( \sigma^{a} \sigma^{e} \sigma^{c}  \sigma^{\rho}\big) \text{tr}\big( \sigma^{b} \sigma^{f} \sigma^{d} \sigma^{\varphi} \big)\\
        &- 4 \sum_{j}\Gamma^{ab,\Lambda}_{i_1 j}(\omega_{1'}+\omega', t, \omega_{1}-\omega') \Gamma^{cd,\Lambda}_{j i_2}(\omega_{2}+\omega',t,-\omega_{2'}+\omega')
        \Pi^{ef,\Lambda}_{jj}(t+\omega',\omega') \text{tr}(\sigma^{b}\sigma^{e}\sigma^{c}\sigma^{f}) \delta_{a\rho} \delta_{d\varphi} \\
        +& 2\Gamma^{ab,\Lambda}_{i_1 i_2}(\omega_{1'}+\omega', t, \omega_{1} - \omega') \Gamma^{cd,\Lambda}_{i_2 i_2}(\omega_{2}+\omega',-\omega_{2'}+\omega',t) \Pi^{ef,\Lambda}_{i_{2}i_{2}}(t+\omega',\omega') \text{tr}\big( \sigma^{d} \sigma^{f} \sigma^{b} \sigma^{e} \sigma^{c} \sigma^{\varphi} \big) \delta_{a\rho}\\
        +& 2\Gamma^{ab,\Lambda}_{i_1 i_1}(\omega_{1'}+\omega',\omega_{1}-\omega',t) \Gamma^{cd,\Lambda}_{i_1 i_2}(\omega_{2}+\omega',t,-\omega_{2'}+\omega')  \Pi^{ef,\Lambda}_{i_{1}i_{1}}(t+\omega',\omega') \text{tr}\big( \sigma^{a} \sigma^{e} \sigma^{c} \sigma^{f} \sigma^{b} \sigma^{\rho}\big) \delta_{d\varphi} \\
        +& \Gamma^{ab,\Lambda}_{i_1 i_2}(\omega_{2'}-\omega', -\omega_{1} - \omega', u) \Gamma^{cd,\Lambda}_{i_1 i_2}(\omega_{2}-\omega',\omega_{1'}+\omega',u) \Pi^{ef,\Lambda}_{i_{2}i_{1}}(u+\omega',\omega')^{*} \text{tr}\big( \sigma^{c} \sigma^{f} \sigma^{a}\sigma^{\rho} \big) \text{tr}\big( \sigma^{b} \sigma^{e} \sigma^{d}\sigma^{\varphi} \big) \Big]
    \end{split}
\end{align}
\end{widetext}
with
\begin{equation}
\Pi^{\rho\varphi,\Lambda}_{i_{1},i_{2}}(\omega_{1},\omega_{2}) = G^{\rho,\Lambda}_{i_{1}}(\omega_{1})\tilde{\mathcal{S}}^{\varphi,\Lambda}_{i_{2}}(\omega_{2})+G^{\varphi,\Lambda}_{i_{2}}(\omega_{2})\tilde{\mathcal{S}}^{\rho,\Lambda}_{i_{1}}(\omega_{1}).
\end{equation}
As already indicated below Eq.~(\ref{eq:BasicFlowEq2}), the single-scale propagator $\mathcal{S}_i^{\rho,\Lambda}(\omega)$ is the negative derivative of $G_i^{\rho,\Lambda}(\omega)$, taking only into account the $\Lambda$-dependence of the cutoff-function but not of the self-energy,
\begin{equation}
\mathcal{S}_i^{\rho,\Lambda}(\omega)=-\frac{d G_i^{\rho,\Lambda}(\omega)}{d \Lambda}\bigg|_{\Sigma^{\Lambda}=\text{const}}.\label{eq:single_scale0}
\end{equation}
Furthermore, the single-scale propagator within Katanin approximation $\tilde{\mathcal{S}}_i^{\rho,\Lambda}(\omega)$ is the full negative derivative of $G_i^{\rho,\Lambda}(\omega)$ which, in the parameterization of Eq.~\eqref{eq:PropagatorPara2}, reads as 
\begin{align}
 &\tilde{\mathcal{S}}_{i}^{\rho,\Lambda}(\omega)=-\frac{d G_{i}^{\rho,\Lambda}(\omega)}{d\Lambda}= \mathcal{S}_i^{\rho,\Lambda}(\omega)   \nonumber\\
 &-\frac{1}{2}\sum_{abc} G_{i}^{a,\Lambda}(\omega) \frac{d \Sigma_{i}^{b,\Lambda}(\omega)}{d\Lambda} G_{i}^{c,\Lambda}(\omega) \text{tr}(\sigma^{a}\sigma^{b}\sigma^{c}\sigma^{\rho}).\label{eq:single_scale}
\end{align}
To make the index structure in Eqs.~(\ref{eq:SEFlowequationImplicit}), (\ref{eq:FlowEqComplete2}), and (\ref{eq:single_scale}) more transparent we have used roman letters $a,b,c,\ldots\in\{0,x,y,z\}$ for internal index summations on the right-hand sides of these equations, while external indices are denoted $\rho,\varphi\in\{0,x,y,z\}$. For an investigation of the general Hamiltonian in Eq.~(\ref{eq:GeneralHamiltonian}), these equations have to be solved for all four components of $\Sigma^\rho$ and 16 components of $\Gamma^{\rho\varphi}$ as well as for all symmetry-inequivalent frequency arguments and site indices. 

\subsection{Discussion of the flow equations and their efficient solution}
The general flow equations in Eqs.~(\ref{eq:SEFlowequationImplicit}) and~(\ref{eq:FlowEqComplete2}) have various characteristic properties that are worth discussing. First, the overall structure of these equations is already well-known from previously investigated PFFRG flow equations for more symmetric spin Hamiltonians. Particularly, the flow equation for the self-energy $\Sigma^{\rho,\Lambda}_i$ on the reference site $i$ has two terms on the right-hand side, the Hartree term [second line of Eq.~(\ref{eq:SEFlowequationImplicit})] and the Fock term [third line of Eq.~(\ref{eq:SEFlowequationImplicit})]. While the Hartree term contains a site summation $\sum_j$, the Fock term is purely local in the sense that {\it only} the reference site $i$ appears as site argument.

The flow equation for the two-particle vertex $\Gamma^{\rho\varphi,\Lambda}_{i_1i_2}$ for the pair of sites $i_1$ and $i_2$ has five terms on the right-hand side [see Eq.~(\ref{eq:FlowEqComplete2})]. In four of these five terms only the sites $i_1$ and $i_2$ appear as site arguments. The remaining term, the so-called random phase approximation (RPA) channel [see fourth line of Eq.~(\ref{eq:FlowEqComplete2})], however, contains a site summation $\sum_j$ such that its evaluation requires more computational effort. We note that a reduced version of the flow equations that only takes into account the Hartee-term in Eq.~(\ref{eq:SEFlowequationImplicit}) and the RPA term in Eq.~(\ref{eq:FlowEqComplete2}) is equivalent to a standard self-consistent spin mean-field theory~\cite{Reuther10,Baez17}. The other terms in the flow equations generate vertex contributions beyond mean field which take into account important effects of quantum fluctuations.

In the RPA channel, the summation over indices $a,\ldots,f\in\{0,x,y,z\}$ generates a smaller number of finite terms than in the other channels of Eq.~(\ref{eq:FlowEqComplete2}), as can be seen as follows. The sum in the RPA channel over indices $a,\ldots,f$ [which is not explicitly carried out in Eq.~(\ref{eq:FlowEqComplete2})] contains a trace over products of Pauli matrices ${\rm tr}(\sigma^{a_1}\cdots \sigma^{a_n})$ and two Kronecker deltas $\delta_{a_1 a_2}$. Among the $4^{n}$ possible index combinations of ${\rm tr}(\sigma^{a_1}\cdots \sigma^{a_n})$ with $a_1,\ldots,a_n\in\{0,x,y,z\}$, only $4^{n-1}$ traces, or one quarter, are non-zero. Similarly, each Kronecker delta $\delta_{a_1 a_2}$ is non-zero in only one quarter of the 16 combinations of indices $a_1$ and $a_2$. Since the RPA channel contains one trace and two Kronecker delta, non-vanishing terms constitute $1/4^3$ of the number of all index combinations $a,\ldots,f$. The other channels each have two traces, or one trace and a Kronecker delta, which reduces the number of non-vanishing terms only by a factor of $1/4^2$ instead.

The number of terms to be evaluated in the RPA channel can be reduced further as follows. The RPA channel contains products of two-particle vertices of the form $\Gamma_{i_1j}^{ab,\Lambda}\Gamma_{j i_2}^{cd,\Lambda}$. One finds that for fixed indices $a$, $b$, $c$, $d$ this product occurs exactly four times in terms that are non-vanishing after the evaluation of traces and Kronecker delta when the remaining summation indices $e$, $f$, as well as the external indices $\rho$, $\varphi$, take all possible values. Thus, one saves computational efforts by evaluating this product, including the site summation over lattice sites $j$, only {\it once} and reusing it in the terms where it is needed. Additionally, one finds for systems with a global U(1) rotation symmetry and no TRS the two-particle vertex flow equation to simplify such (see Sec. \ref{sec:FlowEqSimplifications}) that the product $\Gamma_{i_1j}^{ab,\Lambda}\Gamma_{j i_2}^{cd,\Lambda}$ occurs exactly twice when the remaining indices assume all possible index combinations.

We finally comment on the role of the renormalization group parameter $\Lambda$ in the flow equations. If the regulator is implemented as a sharp cutoff, $G_0^\Lambda=\theta(|\omega|-\Lambda)G_0$, as in our results below, the single scale propagator is proportional to a delta function $\mathcal{S}_i^{\rho,\Lambda}\sim \delta(|\omega|-\Lambda)$ since the $\Lambda$-derivative in Eq.~(\ref{eq:single_scale0}) only acts on the step function $\theta(|\omega|-\Lambda)$. Due to this delta function, the $\omega'$-integration in the non-Katanin terms of Eqs.~(\ref{eq:SEFlowequationImplicit}) and (\ref{eq:FlowEqComplete2}) can be evaluated analytically. However, since most of the numerical effort stems from the Katanin contribution (where the $\omega'$-integration is unavoidable) the benefits of a step-like cutoff are comparatively small.

As already mentioned, the cutoff parameter $\Lambda$ shares some similarities with the temperature $T$. In Ref.~\cite{Iqbal2016} a concrete relation between both quantities has been proposed in the form of the simple proportionality $\Lambda=2T/\pi$. This relation has been derived in the RPA mean-field limit for a sharp frequency cutoff without TRS breaking where the renormalization group equations can be solved exactly and compared to each other when either $\Lambda$ or $T$ take the role of the regulator. Because of its validity only in this special case, the relation $\Lambda=2T/\pi$ should generally be used with caution. Without TRS this proportionality is no longer valid, not even on a mean-field level, but replaced by a more complicated functional relation. This is demonstrated in Appendix~\ref{appendix:AnalyticSolution} for a free spin in a magnetic field and in Appendix~\ref{appendix:Meanfield} for a solution of the Hartree self-consistent equation, both indicating that the association of $\Lambda$ with temperature is not a rigorous correspondence.

\subsection{Flow equation simplifications upon re-introducing symmetries}
\label{sec:FlowEqSimplifications}

\begin{table}[t]
\centering
\begin{tabular}{ |c|c|c|c| }
 \hline  Interaction & TRS & $\Sigma$& $\Gamma$ \\\hline\hline &&&\\[-8pt] 
 Heisenberg & Yes & $\begin{pmatrix} \Sigma^{0} \\ 0 \\ 0 \\ 0\end{pmatrix}$ &$\begin{pmatrix} \Gamma^{\rm d} & 0 & 0 & 0\\
0 & \Gamma^{\rm s} & 0 & 0\\
0 & 0 & \Gamma^{\rm s} & 0\\
0 & 0 & 0 & \Gamma^{\rm s}\end{pmatrix}$ \\[19pt]\hline  &&&\\[-8pt] 
 XYZ & Yes & $\begin{pmatrix} \Sigma^{0} \\ 0 \\ 0 \\ 0\end{pmatrix}$&$\begin{pmatrix} \Gamma^{00} & 0 & 0 & 0\\
0 & \Gamma^{xx} & 0 & 0\\
0 & 0 & \Gamma^{yy} & 0\\
0 & 0 & 0 & \Gamma^{zz}\end{pmatrix}$ \\[19pt]\hline  &&&\\[-8pt] 
 \begin{tabular}{@{}c@{}} \\ U(1) symmetric \\ (about $z$-axis)\end{tabular}  & Yes & $\begin{pmatrix} \Sigma^{0} \\ 0 \\ 0 \\ 0\end{pmatrix}$ &$\begin{pmatrix} \Gamma^{00} & 0 & 0 & \Gamma^{0z}\\
0 & \Gamma^{xx} & \Gamma^{xy} & 0\\
0 & -\Gamma^{xy} & \Gamma^{xx} & 0\\
\Gamma^{z0} & 0 & 0 & \Gamma^{zz}\end{pmatrix}$ \\[19pt]\hline  &&&\\[-8pt] 
 Unconstrained & Yes & $\begin{pmatrix} \Sigma^{0} \\ 0 \\ 0 \\ 0 \end{pmatrix}$ &$\begin{pmatrix} \Gamma^{00} & \Gamma^{0x} & \Gamma^{0y} & \Gamma^{0z}\\
\Gamma^{x0} & \Gamma^{xx} & \Gamma^{xy} & \Gamma^{xz}\\
\Gamma^{y0} & \Gamma^{yx} & \Gamma^{yy} & \Gamma^{yz}\\
\Gamma^{z0} & \Gamma^{zx} & \Gamma^{zy} & \Gamma^{zz}\end{pmatrix}$ \\[19pt]\hline  &&&\\[-8pt] 
\begin{tabular}{@{}c@{}} \\ U(1) symmetric \\ (about $z$-axis)\end{tabular} & No & $\begin{pmatrix} \Sigma^{0} \\ 0 \\ 0 \\ \Sigma^{z}\end{pmatrix}$ &$\begin{pmatrix} \Gamma^{00} & 0 & 0 & \Gamma^{0z}\\
0 & \Gamma^{xx} & \Gamma^{xy} & 0\\
0 & -\Gamma^{xy} & \Gamma^{xx} & 0\\
\Gamma^{z0} & 0 & 0 & \Gamma^{zz}\end{pmatrix}$ \\[19pt]\hline  &&&\\[-8pt] 
 Unconstrained & No & $\begin{pmatrix} \Sigma^{0} \\ \Sigma^{x} \\ \Sigma^{y} \\ \Sigma^{z}\end{pmatrix}$  &$\begin{pmatrix} \Gamma^{00} & \Gamma^{0x} & \Gamma^{0y} & \Gamma^{0z}\\
\Gamma^{x0} & \Gamma^{xx} & \Gamma^{xy} & \Gamma^{xz}\\
\Gamma^{y0} & \Gamma^{yx} & \Gamma^{yy} & \Gamma^{yz}\\
\Gamma^{z0} & \Gamma^{zx} & \Gamma^{zy} & \Gamma^{zz}\end{pmatrix}$  \\[19pt]\hline
\end{tabular}
\caption{Finite spin components of the self-energy $\Sigma^\rho$ (third column) and the two-particle vertex $\Gamma^{\rho\varphi}$ (fourth column) for $\rho,\varphi\in\{0,x,y,z\}$ for different types of spin models with distinct symmetries. The considered spin models are characterized by their types of two-body spin interactions or by their spin rotation symmetries (first column), and whether they possess TRS (second column). An XYZ interaction is of the form $J_x \hat{S}_i^x \hat{S}_j^x+J_y \hat{S}_i^y \hat{S}_j^y+J_z \hat{S}_i^z \hat{S}_j^z$. Frequency and site arguments are omitted for brevity. Components which are equal by symmetry, are labeled identically.}
\label{tab:VertexStructure}
\end{table}

\begin{table}[t]
\centering
\begin{tabular}{ |c|c|c| }
 \hline Interactions & TRS & \begin{tabular}{@{}c@{}}Relative number of \\ products $\Gamma^{ab}\Gamma^{cd}$\end{tabular}\\\hline\hline
 Heisenberg & Yes & 1 \\[4pt]
 XYZ & Yes & 2 \\[4pt]
 U(1) symmetric & Yes  & 6 \\[4pt]
 Unconstrained & Yes & 32 \\[4pt]
 U(1) symmetric & No & 10 \\[4pt]
 Unconstrained & No & 128 \\ \hline
\end{tabular}
\caption{Relative number of independent and finite two-particle vertex products $\Gamma^{ab} \Gamma^{cd}$ in the RPA channel for unconstrained indices $a,b,c,d \in \{0,x,y,z\}$ (frequency and site arguments are kept implicit and differ between $\Gamma^{ab}$ and $\Gamma^{cd}$). The same type of models as in Table~\ref{tab:VertexStructure} are considered, i.e., systems with and without TRS.}
\label{tab:RPA2PVertexProducts}
\end{table}

The flow equations in Eqs.~\eqref{eq:SEFlowequationImplicit} and \eqref{eq:FlowEqComplete2} do not assume any global spin rotation symmetry or TRS. If one re-introduces these symmetries, the flow equations simplify. For example, in the presence of TRS the components of $\Sigma^\mu$ and $G^\mu$ with $\mu\in\{x,y,z\}$ vanish \cite{Buessen19}.
While TRS allows for each of the 16 two-particle vertex function components $\Gamma^{\rho\varphi}$ to be finite and independent, continuous spin rotation symmetries can enforce components to vanish or become dependent on each other. To showcase the complexity of the PFFRG flow equations for different types of imposed symmetries, Table~\ref{tab:VertexStructure} displays the spin structure of the self-energy and two-particle vertex for different types of spin Hamiltonians with two-body spin interactions in the presence and absence of TRS.
The often considered case of a Heisenberg model without magnetic fields only results in finite two-particle vertex components $\Gamma^{00}$ and $\Gamma^{xx}=\Gamma^{yy}=\Gamma^{zz}$, usually referred to as $\Gamma^{\rm d}$ and $\Gamma^{\rm s}$, respectively~\cite{Reuther10}. In comparison, a model with global U(1) spin rotation symmetry (with or without TRS) produces six independent two-particle vertex components~\cite{Hering17}. Note that, apart from XXZ couplings $J_{\perp}( \hat{S}_i^x \hat{S}_j^x+\hat{S}_i^y \hat{S}_j^y)+J_z \hat{S}_i^z \hat{S}_j^z$, a spin system with global U(1) spin rotation symmetry around the $z$ axis can also contain a Dzyaloshinskii–Moriya term of the form $(\hat{\bm S}_i\times \hat{\bm S}_j)^z$. XYZ models with coupling terms $J_x \hat{S}_i^x \hat{S}_j^x+J_y \hat{S}_i^y \hat{S}_j^y+J_z \hat{S}_i^z \hat{S}_j^z$ which (apart from special cases) do not exhibit continuous spin rotation symmetries reduce the number of independent finite two-particle vertex components to four if TRS is intact.

It follows from the reduced vertex spin structures of Table~\ref{tab:VertexStructure}, that simplified flow equations of the above mentioned models are straightforwardly obtained by restricting the allowed values for summation indices $a,\ldots,f$, as well as for $\rho$ and $\varphi$, in Eqs.~\eqref{eq:SEFlowequationImplicit} and~\eqref{eq:FlowEqComplete2}.

The symmetry-imposed simplification of vertex structures is reflected in the number of independent and finite two-particle vertex products $\Gamma^{ab}\Gamma^{cd}$ which appear in the RPA channel for unconstrained indices $a,b,c,d \in \{0,x,y,z\}$, as summarized in Table \ref{tab:RPA2PVertexProducts}. As a trend, excluding the XYZ model, the number of terms increases for each broken continuous spin rotation symmetry. However, we point out that a U(1) symmetric model with broken TRS still generates fewer independent terms in the RPA channel than a model with TRS and without any continuous spin rotation symmetry. Since the numerical resources needed to evaluate the flow equations are mostly determined by the RPA channel, the relative numbers of Table \ref{tab:RPA2PVertexProducts} reflect the overall effort for solving the flow equations. Note, however, that broken TRS removes the two-particle vertex frequency symmetry in Eq. \eqref{eq:TRSFrequencySymmetry} as well, doubling the effort of evaluating the flow equations.

\subsection{Initial conditions}
Finally, to complete the discussion of the PFFRG flow equations for TRS-broken systems, we specify the initial conditions at $\Lambda\rightarrow\infty$. In this limit only the bare parameters of the spin Hamiltonian in Eq.~(\ref{eq:GeneralHamiltonian}) remain~\cite{Metzner12, Mueller23}, yielding the initial conditions for the self-energy,
\begin{equation}
    \Sigma^{\mu,\Lambda \rightarrow \infty}_{i}(\omega) = - \frac{1}{2} h^{\mu}_{i},\label{eq:ini_self}
\end{equation}
and for the two-particle vertex
\begin{equation}
    \Gamma^{\mu\nu,\Lambda \rightarrow \infty}_{ij}(s,t,u) = \frac{1}{4} J^{\mu\nu}_{ij},\label{eq:ini_two_particle}
\end{equation}
where $\mu,\nu \in \{x,y, z\}$. Note that the factors $\frac{1}{2}$ and $\frac{1}{4}$ in Eqs.~(\ref{eq:ini_self}) and (\ref{eq:ini_two_particle}), respectively, are due to the factor $\frac{1}{2}$ in the fermion mapping, see Eq.~(\ref{eq:pseudoFermionMapping}).

\subsection{\label{sec:Observables} Physical observables}
The vertex functions $\Sigma_i^{\rho,\Lambda}$ and $\Gamma_{i_1i_2}^{\rho\varphi,\Lambda}$ obtained from solving the PFFRG flow equations are no physical observables themselves but can be straightforwardly used to calculate those quantities. A standard physical outcome in previous PFFRG implementations (and in our extension) are static and equal time spin-spin correlation functions $\chi^{\mu\nu}_{ij}(\omega=0)$ and $\chi^{\mu\nu}_{ij}(\tau=0)$, respectively, defined by
\begin{equation}
    \chi^{\mu\nu}_{ij}(\omega) = \int_{0}^{\infty} d\tau e^{i\omega\tau} \big( \langle\hat{S}^{\mu}_{i}(\tau)\hat{S}^{\nu}_{j}(0) \rangle - \langle \hat{S}^{\mu}_{i}(\tau) \rangle \langle \hat{S}^{\nu}_{j}(0) \rangle \big)
\end{equation}
and
\begin{equation}
\chi^{\mu\nu}_{ij}(\tau=0)=\frac{1}{2\pi}\int d\omega \chi^{\mu\nu}_{ij}(\omega)=\langle\hat{S}_i^\mu\hat{S}_j^\nu\rangle-\langle \hat{S}_i^\mu\rangle \langle \hat{S}_j^\nu\rangle
\end{equation}
with $\mu,\nu \in \{x,y,z\}$. Note that $\omega\in \mathbb{R}$ is the imaginary part of a purely imaginary frequency such that $\chi^{\mu\nu}_{ij}(\omega\neq0)$ does not directly correspond to a physical observable.
After mapping the spin operators onto pseudo-fermions, one can express $\chi^{\mu\nu,\Lambda}_{ij}(\omega)$ (now equipped with an explicit $\Lambda$-dependence) in terms of the vertex functions as follows
\begin{widetext}
\begin{align}
\label{eq:SpinCorrelation}
    \chi^{\mu\nu,\Lambda}_{ij}(\omega)&=-\frac{1}{8\pi} \delta_{ij} \sum_{ab} \int d\omega'\big( G^{a,\Lambda}_{i}(\omega') G^{b,\Lambda}_{i}(\omega+\omega')  \text{tr}(\sigma^{\mu}\sigma^{a}\sigma^{\nu}\sigma^{b})  \big)\nonumber\\ 
    &-\frac{1}{16\pi^{2}} \sum_{abcdef} \int d\omega' d\omega'' G^{a,\Lambda}_{i}(\omega')G^{b,\Lambda}_{j}(\omega+\omega'')G^{c,\Lambda}_{i}(\omega+\omega')G^{d,\Lambda}_{j}(\omega'')\big(\Gamma^{ef,\Lambda}_{ij}(\omega+\omega'+\omega'',\omega,\omega'-\omega'')\times\nonumber\\ &\text{tr}(\sigma^{\mu}\sigma^{c}\sigma^{e}\sigma^{a}) \text{tr}(\sigma^{\nu}\sigma^{d}\sigma^{f}\sigma^{b})  - \delta_{ij} \Gamma^{ef,\Lambda}_{ij}(\omega+\omega'+\omega'',\omega'-\omega'',\omega) \text{tr}(\sigma^{\mu}\sigma^{c}\sigma^{e}\sigma^{b}\sigma^{\nu}\sigma^{d}\sigma^{f}\sigma^{a}) \big).
\end{align}
\end{widetext}
In case TRS is broken one can also compute a finite magnetization $M^{\mu}_{i} = \langle \hat{S}^{\mu}_{i}\rangle$ from the single particle Green function,
\begin{align}
\label{eq:Magnetization}
M^{\mu,\Lambda}_{i} =& \frac{1}{2\pi}\int d\omega g^{\mu,\Lambda}_{i}(\omega).
\end{align}

\section{\label{sec:Application}Results: Magnetizations of 2D spin models}

We now present results of the TRS-breaking PFFRG formalism developed in the last sections by treating spin models in the presence of finite magnetic fields. In our exploratory investigations, we focus on well-studied 2D spin models on square, honeycomb and triangular lattices and compare our results with existing literature. This will expose various advantages and disadvantages of the method which may serve as a guide for future applications.

Below, we pursue two different strategies of including magnetic fields, that are reflected in the following Hamiltonian,
\begin{equation}
\label{eq:HeisenbergInBField}
\hat{\mathcal{H}} = J \sum_{\langle ij \rangle} \hat{\bm S}_{i} \cdot \hat{\bm S}_{j} - \delta \sum_{i} {\bm n}_{i} \cdot \hat{\bm S}^\mu_{i} - h \sum_{i} \hat{S}^{z}_{i}.
\end{equation}
The first term contains the two-body spin interactions, which, for the systems considered here, are of nearest neighbor Heisenberg type. The second term is a small seed field $\delta\ll|J|$, which breaks the system's symmetries (that would otherwise be spontaneously broken) already on the Hamiltonian level such that the renormalization group flows can be continued into long-range ordered phases. Unless stated otherwise, the orientations of the vectors ${\bm n}_i$ (which we normalize as $|{\bm n}_i|=1$) are fixed according to the expected long-range orders. However, as we will demonstrate below for the Heisenberg antiferromagnet on the triangular lattice, the correct type of order is even generated if these vectors deviate from it, as long as the perturbation breaks the otherwise spontaneously broken symmetries. We apply small seed fields to regularize renormalization group flows (Sec.~\ref{sec:flow_regularization}) and to compute the magnetizations for different long-range ordered systems in the limit $\delta\rightarrow 0$ (Sec.~\ref{sec:zerofield}).

The third term in Eq.~(\ref{eq:HeisenbergInBField}) corresponds to a uniform external magnetic field, which (without loss of generality) we orient along the $z$-axis. We consider finite $h>0$ to investigate the magnetization process of a spin system up to saturation. If a system's ground state at finite $h$ exhibits a spontaneous breaking of spin rotation or lattice symmetries not already captured by this third term, it is necessary to introduce additional small fields $\delta$, such that the renormalization group flow can be continued into this symmetry broken phase. An example for this type of application will be presented for the square lattice Heisenberg antiferromagnet (Sec.~\ref{sec:magnetization_curve}).

The PFFRG is applied with the following specifications. The two-particle vertex is computed on a frequency grid with $92$ points ($76$ points for triangular lattice models) for each frequency $s$, $t$, $u$, while the self-energy is calculated for $2000$ frequencies. Our frequency grids have an exponential distribution of grid points and we use a sharp frequency cutoff $G_0^\Lambda=\theta(|\omega|-\Lambda)G_0$. To improve the frequency resolution of vertex functions, it is beneficial to implement the integrations with an adaptive frequency grid~\cite{Ritter22, Kiese22, Thoenniss20}, which, however, we postpone to future work. Unless stated otherwise, correlations of square and triangular lattice models are neglected beyond distances of $L=5$ nearest neighbor spacings. An explicit embedded Runge-Kutta (2, 3) method with adaptive step size is used to solve the flow equations~\cite{GSL09}.

\begin{figure*}[t!]
    \centering
    \vspace{5mm}
\begin{subfigure}[t]{0.46\textwidth}
        \centering
        \begin{overpic}[width = \textwidth]{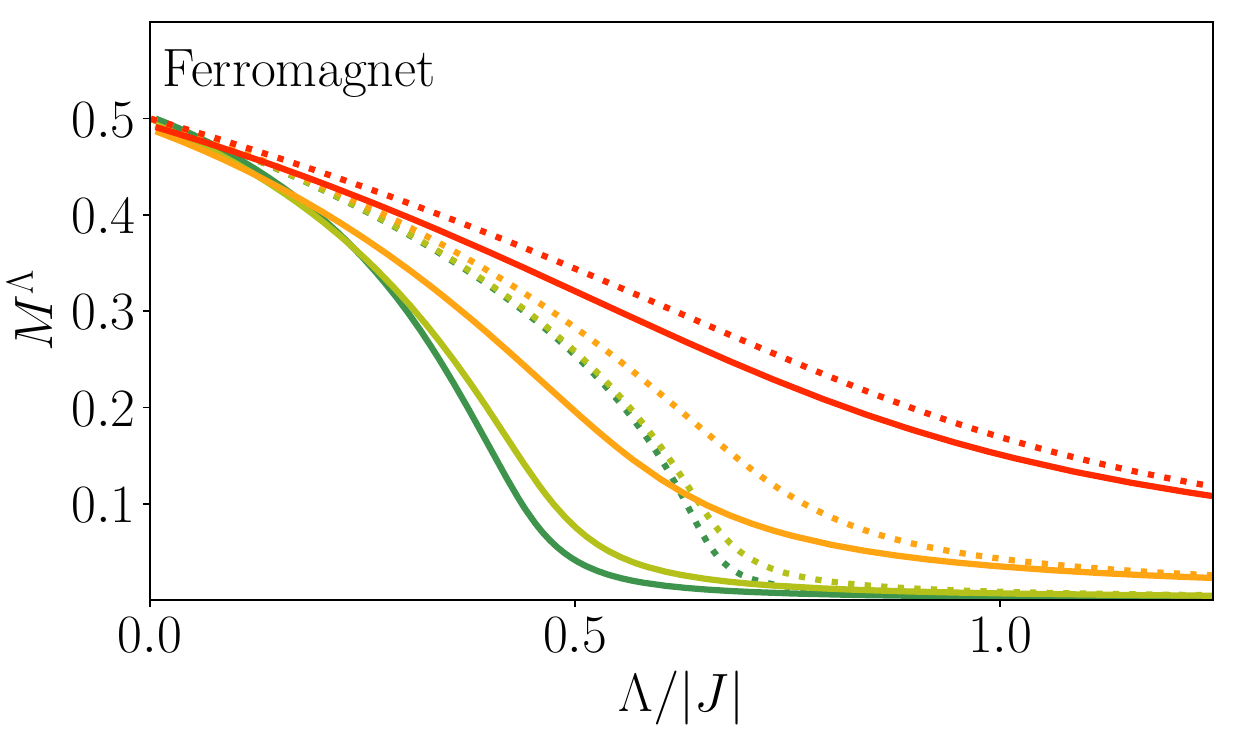}
            \put(12,60){\includegraphics[scale=0.9]{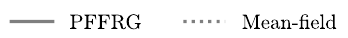}}
            \put(70,29){\includegraphics[scale=.2]{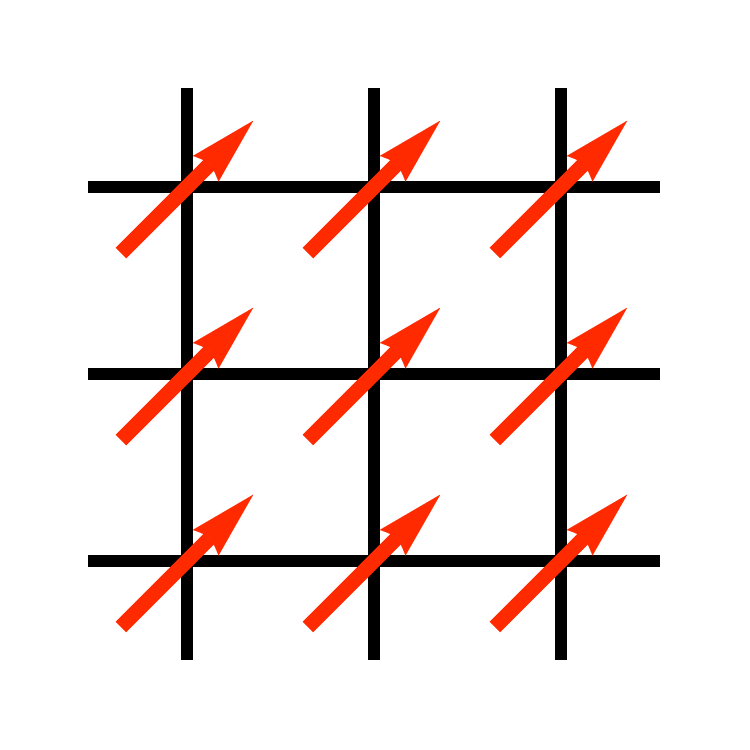}}
            \put(-4,62){(a)}
        \end{overpic}
\end{subfigure}
\hspace{5mm}
\begin{subfigure}[t]{0.46\textwidth}
        \centering
        \begin{overpic}[width = \textwidth]{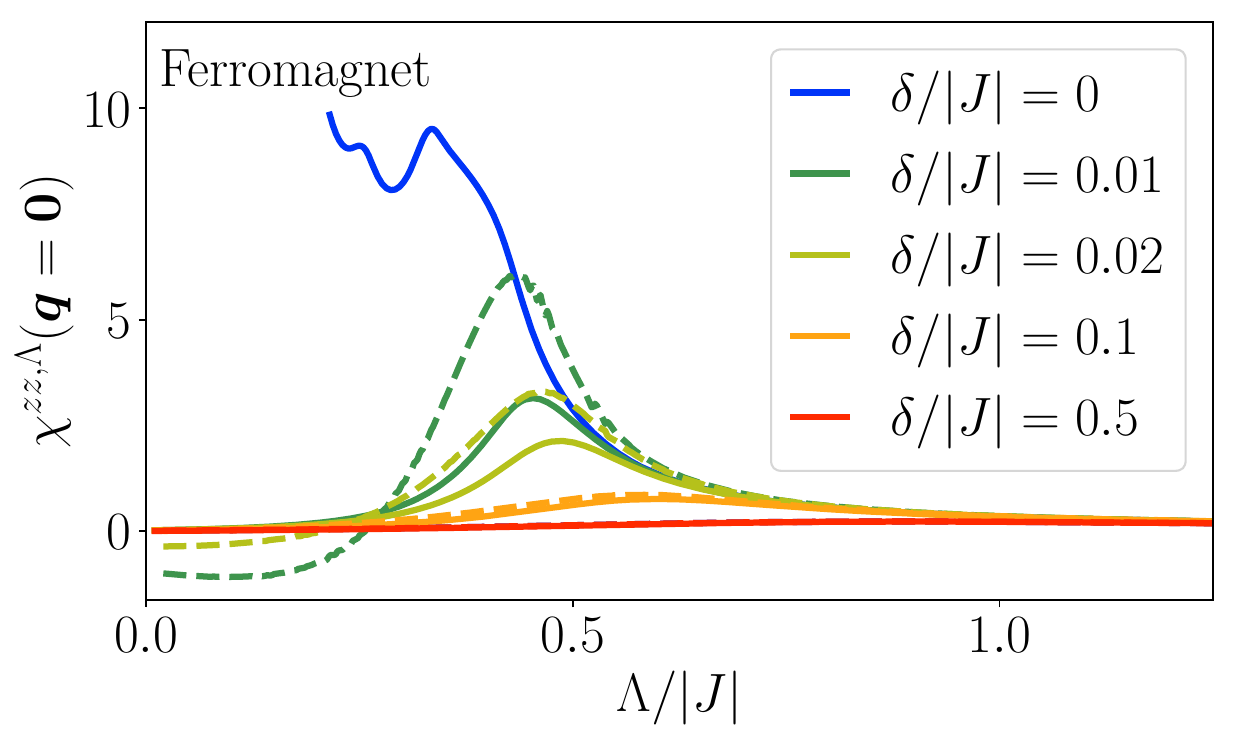}
            \put(12,60){\includegraphics[scale=0.9]{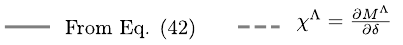}}
            \put(-4,62){(b)}
        \end{overpic}
\end{subfigure}
\begin{subfigure}[t]{0.46\textwidth}
        \centering
        \begin{overpic}[width = \textwidth]{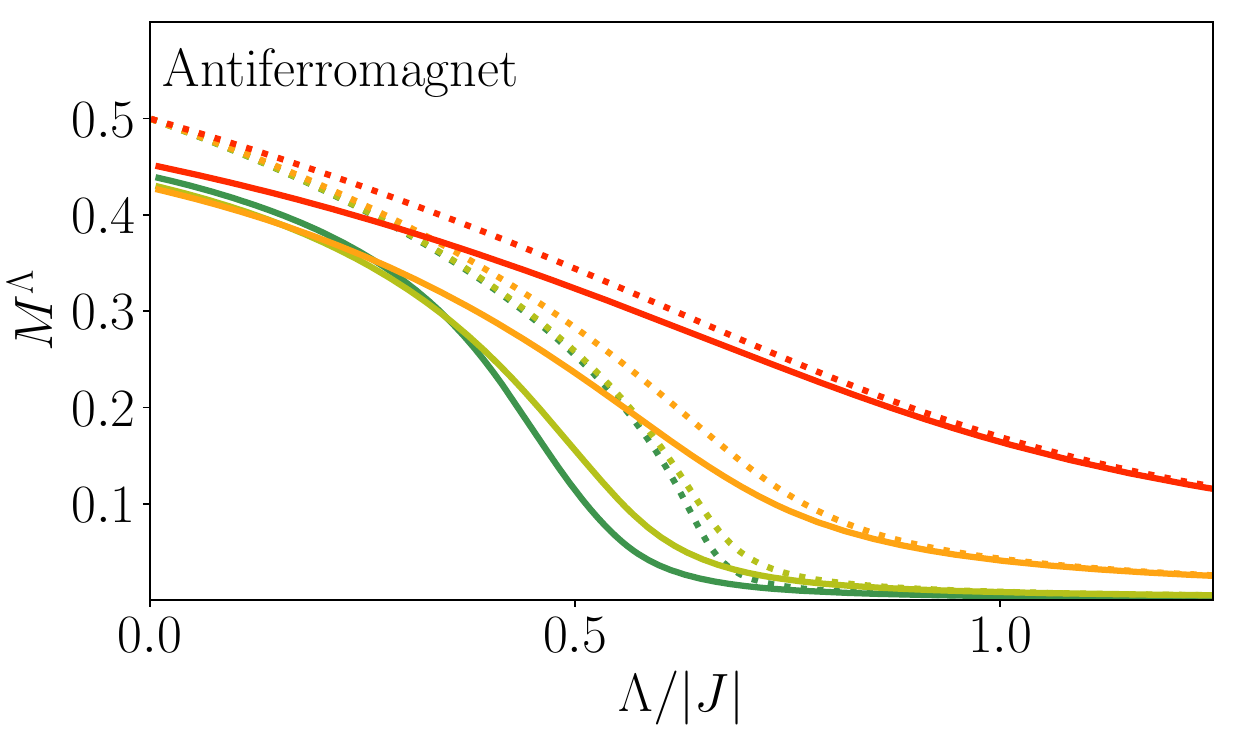}
            \put(70,29){\includegraphics[scale=.2]{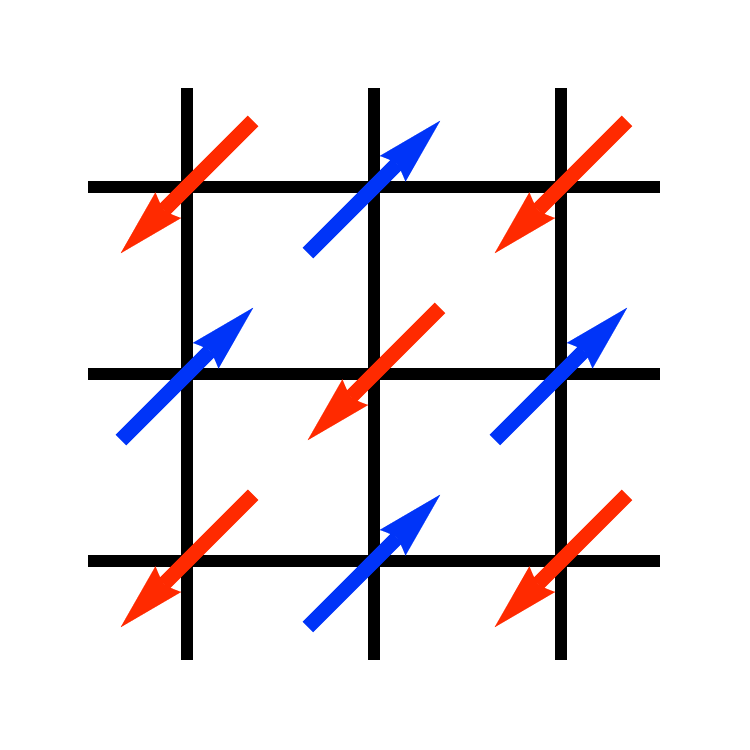}}
            \put(-4,56){(c)}
        \end{overpic}
\end{subfigure}
\hspace{5mm}
\begin{subfigure}[t]{0.46\textwidth}
        \centering
        \begin{overpic}[width = \textwidth]{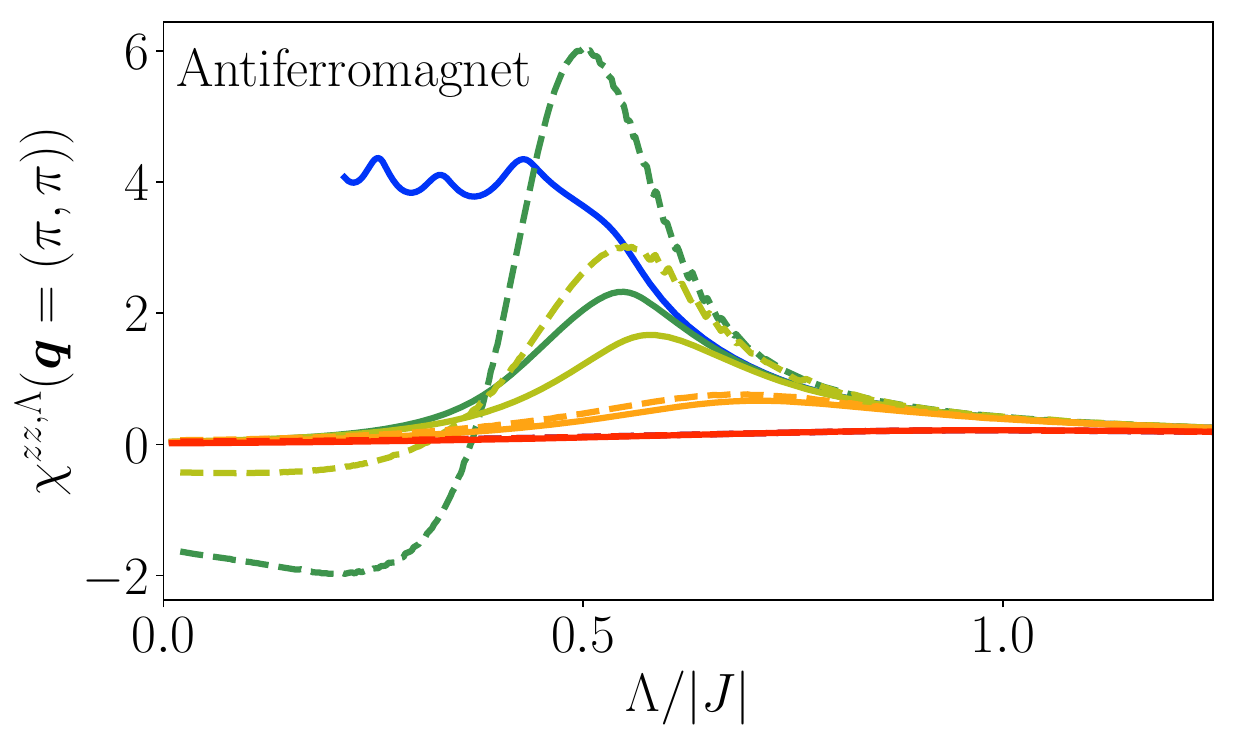}
            \put(-4,56){(d)}
        \end{overpic}
\end{subfigure}
        \caption{(a), (c) Magnetization $M^\Lambda=|{\bm M}_i^\Lambda|$ as a function of $\Lambda$ for the (a) ferromagnetic and (c) antiferromagnetic square lattice Heisenberg model. Full (dotted) lines correspond to PFFRG (mean-field) results. Different colors correspond to different strengths of seed fields ($\delta/|J|=0$, $0.01$, $0.02$, $0.1$ and $0.5$), see legend in subfigure (b) which applies to all subfigures. Seed fields are oriented parallel or antiparallel to the $z$-axis according to the insets. (b), (d) Longitudinal order parameter susceptibilities $\chi^{zz,\Lambda}({\bm q})$ as a function of $\Lambda$ for the (b) ferromagnetic and (d) antiferromagnetic square lattice Heisenberg model where ${\bm q}={\bm 0}$ and ${\bm q}=(\pi,\pi)$, respectively. Full lines are PFFRG results derived from Eq.~(\ref{eq:SpinCorrelation}) while dashed lines are derived from $\chi^\Lambda=\partial M^\Lambda/\partial \delta$. In the latter approach, $\delta$-derivatives are approximated by the variation of $M^\Lambda$ when $\delta$ is increased by $10\%$ from the value stated in the legend.}
        \label{fig:PerturbedModels}
\end{figure*}

\subsection{Flow regularization by small seed fields $\delta$}\label{sec:flow_regularization}
The successful regularisation of susceptibility flow breakdowns by finite fields $\delta$ (at $h=0$) is demonstrated in Fig.~\ref{fig:PerturbedModels} for both the ferromagnetic and antiferromagnetic Heisenberg models on the square lattice. The distinct differences of susceptibility flows with and without seed fields are shown in Fig.~\ref{fig:PerturbedModels}(b) and (d), see the full lines, computed from Eq.~(\ref{eq:SpinCorrelation}). For $\delta=0$ the onset of magnetic long-range order is indicated by an increase of the respective order parameter susceptibilities $\chi^{zz,\Lambda}({\bm q}={\bm 0})$ and $\chi^{zz,\Lambda}({\bm q}=(\pi,\pi))$ followed by a characteristic kink. At cutoffs below this breakdown feature the susceptibility no longer has any physical meaning (as a result of the frequency discretization and sharp cutoff function, the susceptibility flow is strongly oscillatory). Once a small seed field $\delta/|J| \geq 0.01$ is switched on, the kink and the subsequent oscillatory behavior are replaced by a smooth peak and the renormalization group flow can be continued down to $\Lambda\rightarrow 0$. 
Note that high numerical stability is required because flow instabilities may reappear when choosing the seed field $\delta$ too small. Such tendency lies in contrast to the objective to choose the seed field as small as possible to avoid the introduced energetic bias which favors spins towards an orientation along its site-dependent magnetic field.
In our numerical implementation, a flow regularization was always successful when applying seed fields of size $\delta/|J| \geq 0.02$.
With increasing $\delta$, the peaks in Fig.~\ref{fig:PerturbedModels}(b) and (d) are suppressed, broadened and shifted to larger $\Lambda$. If $\Lambda$ is interpreted as a temperature, these observations qualitatively resemble the expected response of a finite temperature magnetic phase transition to a magnetic field. Note, however, that the existence of a finite critical scale $\Lambda_c$ is an artifact from the truncation of renormalization group equations~\cite{Mueller23}, since Mermin-Wagner theorem forbids a finite temperature (or finite $\Lambda$) transition in two-dimensional Heisenberg models.

The magnetizations $M^\Lambda=|{\bm M}^\Lambda_i|=|\langle {\bm S}_i\rangle^\Lambda|$ as a function of $\Lambda$ [see Eq.~(\ref{eq:Magnetization})] for the ferromagnetic and antiferromagnetic square lattice Heisenberg models in Fig.~\ref{fig:PerturbedModels}(a) and (c), respectively, also show qualitatively correct behavior. For small $\delta$ the onset of magnetic order is indicated by a sudden increase of $M^\Lambda$ at a finite value of $\Lambda$ which approximately coincides with the position of the susceptibility peaks in Fig.~\ref{fig:PerturbedModels}(b) and (d). With increasing $\delta$, the curvature of the magnetization around this upturn decreases. For comparison, Fig.~\ref{fig:PerturbedModels}(a) and (c) also shows the results of a bare self-consistent mean-field calculation (which is equivalent to only taking into account the Hartree and RPA channels in the PFFRG flow equations), see dotted lines. Details can be found in Appendix~\ref{appendix:Meanfield}. The onset of magnetization in the full PFFRG occurs at smaller values of $\Lambda$ compared to the bare mean field calculation, which is due to the quantum fluctuations contained in the PFFRG but not in the mean field approach. Furthermore, for the antiferromagnetic model the magnetization at $\Lambda\rightarrow0$ is reduced from the mean-field value $M=\frac{1}{2}$, which will be discussed is more detail in Sec.~\ref{sec:zerofield}.

One generally expects that the magnetizations $M^\Lambda$ increase monotonically with increasing seed fields $\delta/|J|$. However, Fig.~\ref{fig:PerturbedModels}(a) and (c) reveals that this is not always the case at small $\Lambda$. We interpret this unphysical observation as a numerical artifact caused by an oscillatory behavior of the frequency dependence of vertex functions at small $\Lambda$. We therefore believe that this problem can be cured by the improved numerical implementation described in Ref.~\cite{Kiese22}.

The accessibility of finite magnetizations allows for an alternative computation of susceptibilities, namely by directly taking the derivative of the magnetization with respect to the field strength, $\chi^\Lambda=\partial M^\Lambda/\partial \delta$. The two ways of calculating the susceptibility, i.e., either via Eq.~(\ref{eq:SpinCorrelation}) or via the derivative of the magnetization are compared in Fig.~\ref{fig:PerturbedModels}(b) and (d), see full and dashed lines, respectively. In the current one-loop plus Katanin truncation the two methods are generally not identical. Diagrammatic approximations which have the property of identical response functions in both approaches have been investigated systematically~\cite{Baym61,Baym62}, and are known as conserving approximations with the property of satisfying conservation laws. Even though the one-loop plus Katanin PFFRG is not such a conserving approximation the degree of agreement between both approaches can be taken as a measure for the quality of our numerical outcomes. While the agreement becomes better upon increasing both, $\delta$ and $\Lambda$, the largest deviations are observed for small $\delta/|J|$ in the $\Lambda$ region around the susceptibility peak. This regime close to a phase transition is well-known to pose challenges for the PFFRG, due to the intricate competition between ordering tendencies and fluctuations~\cite{Mueller23}. Furthermore, the unphysical negative susceptibilities $\chi^\Lambda=\partial M^\Lambda/\partial \delta$ at small $\Lambda$ and $\delta$ are a consequence of the aforementioned non-monotonic behavior of the magnetization in $\delta$.

\begin{figure}[t]
\begin{subfigure}[t]{0.28\columnwidth}
        \centering
        \hspace{-1.8cm}
        \begin{overpic}[width = \textwidth]{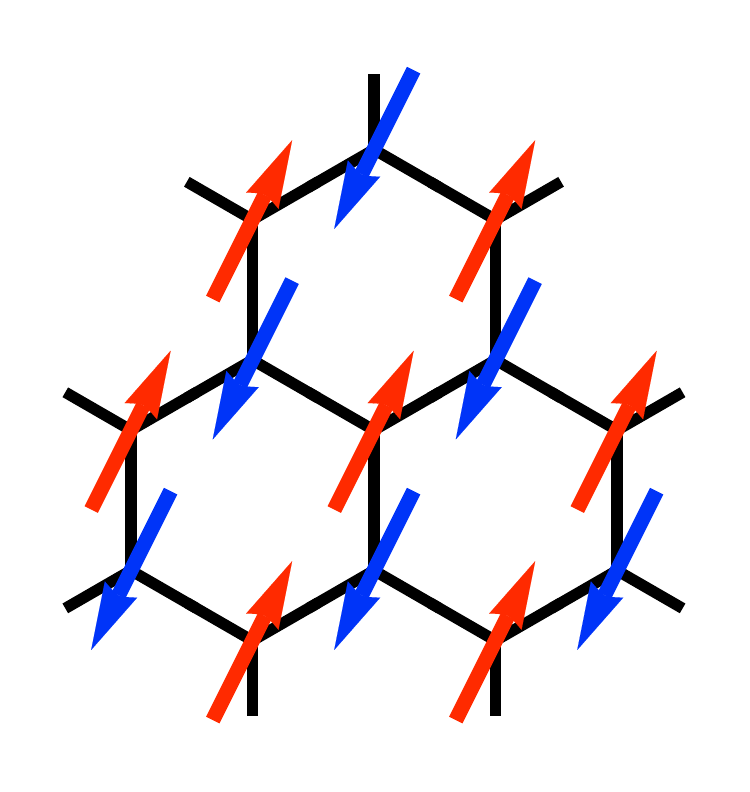}
        \put(-25,80){(a)}
        \end{overpic}
\end{subfigure}
\vspace{-3mm}
\hspace{8mm}
\begin{subfigure}[t]{0.28\columnwidth}
        \centering
        \hspace{-1cm}
        \begin{overpic}[width = \textwidth]{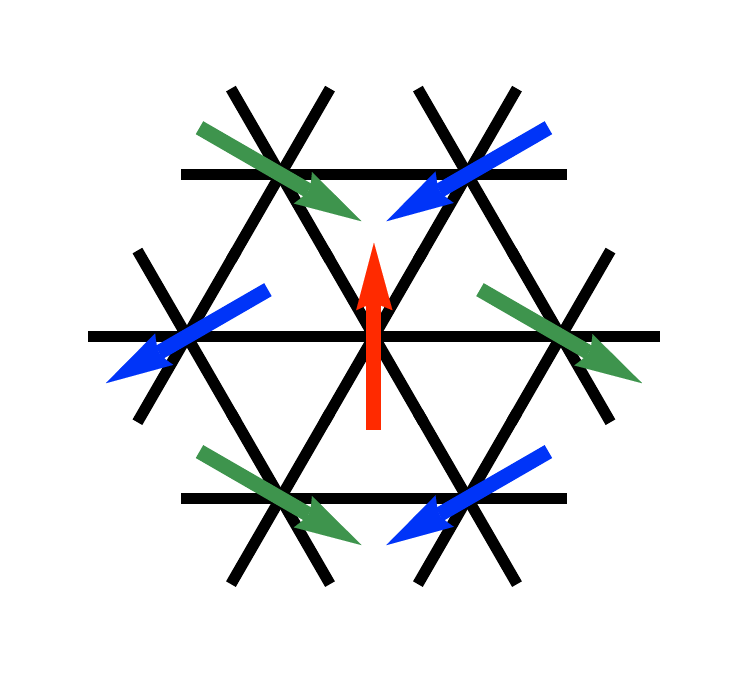}
        \put(-10,85){(b)}
        \end{overpic}
\end{subfigure}\vspace{2mm}
    \begin{subfigure}[t]{\columnwidth}
        \centering
        \begin{overpic}[width = \textwidth]{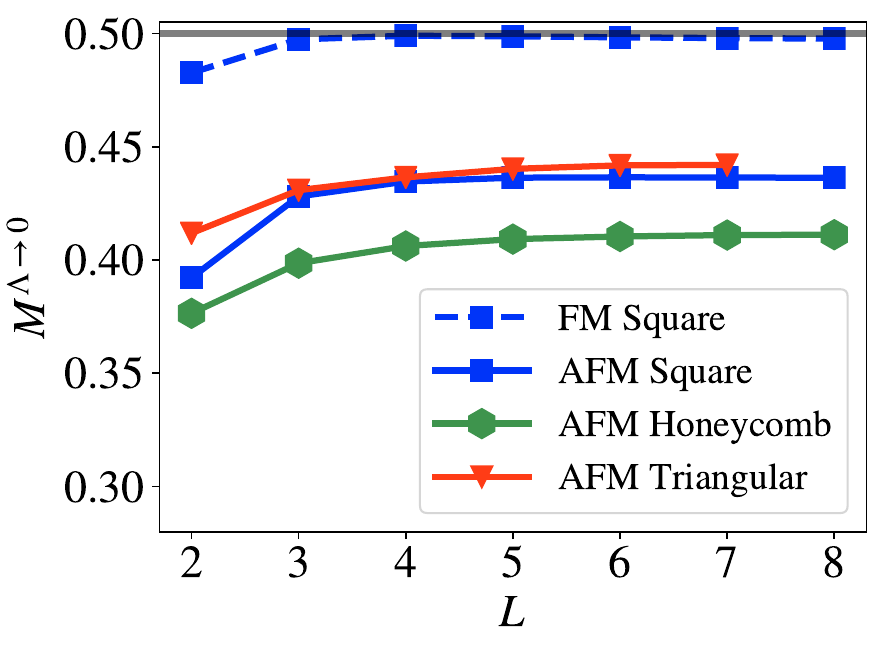}
        \put(-1.4,68){(c)}
        \end{overpic}
    \end{subfigure}
            \vspace{-6mm}
        \caption{(a) N\'eel order on the honeycomb lattice. (b) Three sublattice $120\degree$ N\'eel order. Different sublattices are distinguished by different colors of the spins. (c) Magnetization $M^{\Lambda\rightarrow0}$ in the small-$\Lambda$ limit as a function of maximum correlation distance $L$, for the ferromagnetic Heisenberg model on the square lattice as well as for the antiferromagnetic Heisenberg models on the square, honeycomb and triangular lattices. Here, $L$ is defined as the maximal distance of spin-spin correlations (in units of the lattice constant), while longer distance spin-spin correlations are treated as zero. The magnetizations shown are obtained at cutoffs $\frac{\Lambda}{J}=0.02$, except for the square lattice ferromagnet where $\frac{\Lambda}{J}=0.01$ is used. Seed fields are of size $\frac{\delta}{J}=0.01$ on the square lattice, and of size $\frac{\delta}{J}=0.02$ on the honeycomb and triangular lattices. 
        }
        \label{fig:staggeredMagnetization}
\end{figure}

\subsection{Zero-field magnetizations}\label{sec:zerofield}
Zero-field magnetizations at $T=0$ that are reduced from saturation $M=\frac{1}{2}$ due to quantum fluctuations are a characteristic property of quantum spin systems. In PFFRG they are obtained from $M^\Lambda$ in the limits $\delta\rightarrow0$ and $\Lambda\rightarrow0$. The ferromagnetic square lattice Heisenberg model [see Fig.~\ref{fig:PerturbedModels}(a)] shows a saturated magnetization $M=\frac{1}{2}$ in this limit, demonstrating that PFFRG correctly captures the absence of quantum fluctuations in the ferromagnatic ground state. In contrast, for the antiferromagnetic Heisenberg model on the square lattice [see Fig.~\ref{fig:PerturbedModels}(c)], quantum fluctuations are observed to reduce the magnetization at $\delta\rightarrow0$ and $\Lambda\rightarrow0$ compared to saturation.

To investigate the reduction of the magnetization due to quantum fluctuations more systematically, we show in Fig.~\ref{fig:staggeredMagnetization}(c) the magnetizations as a function of maximum correlation distance $L$ for a variety of different 2D Heisenberg models. In addition to the square lattice, we also present results for the nearest neighbor Heisenberg models on the honeycomb and triangular lattices. The honeycomb lattice (like the square lattice) is bipartite and exhibits a collinear N\'eel ground state in the antiferromagnetic case [see Fig.~\ref{fig:staggeredMagnetization}(a)]. The corresponding seed field $\sim{\bm n}_i$, hence, maintains a U(1) spin rotation symmetry around the magnetization axis. On the other hand, the triangular lattice antiferromagnet is known to feature non-collinear $120\degree$ ground state order~\cite{White07,huang23,Capriotti99} depicted in Fig.~\ref{fig:staggeredMagnetization}(b). Consequently, the corresponding seed fields break all continuous spin rotation symmetries. This increases the computational effort and we have, consequently, refrained from performing a calculation with $L=8$ for this system, see Fig.~\ref{fig:staggeredMagnetization}(c).

\begin{figure}[t]
    \centering
    \begin{subfigure}[t]{0.28\columnwidth}
        \centering
        \hspace{-1.8cm}
        \begin{overpic}[width = \textwidth]{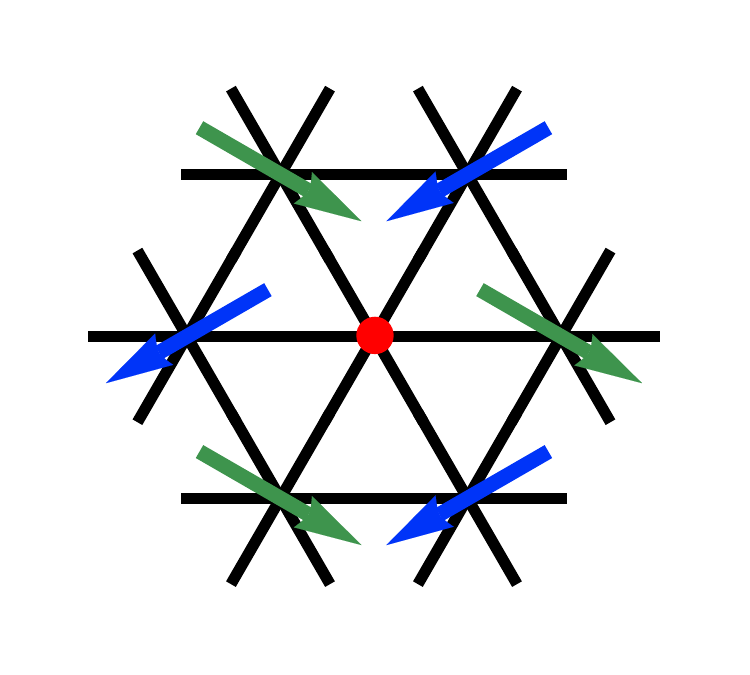}
        \put(-20,80){(a)}
        \end{overpic}
\end{subfigure}
\hspace{8mm}
\begin{subfigure}[t]{0.28\columnwidth}
        \centering
        \hspace{-1cm}
        \begin{overpic}[width = \textwidth]{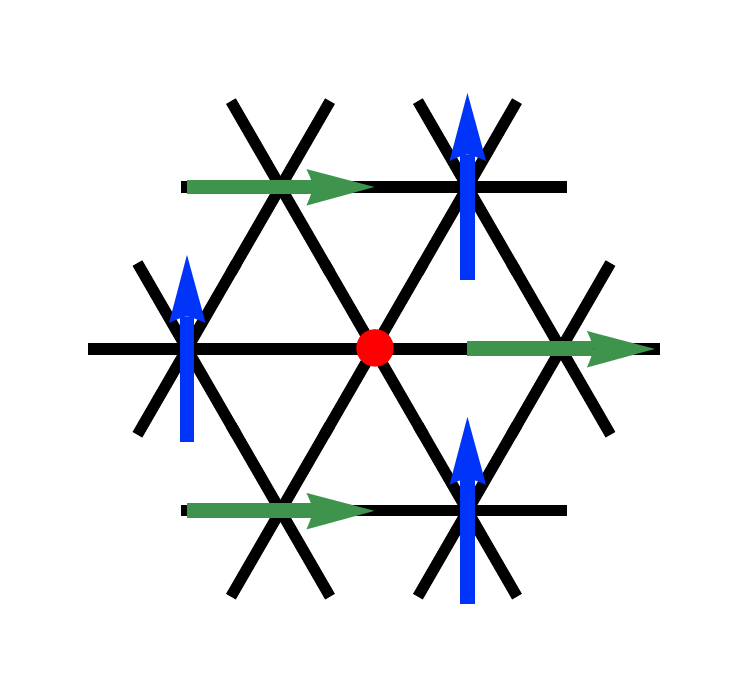}
        \put(-20,80){(b)}
        \end{overpic}
\end{subfigure}
    \begin{subfigure}[t]{1\columnwidth}
        \centering
        \begin{overpic}[width = 0.9\textwidth]{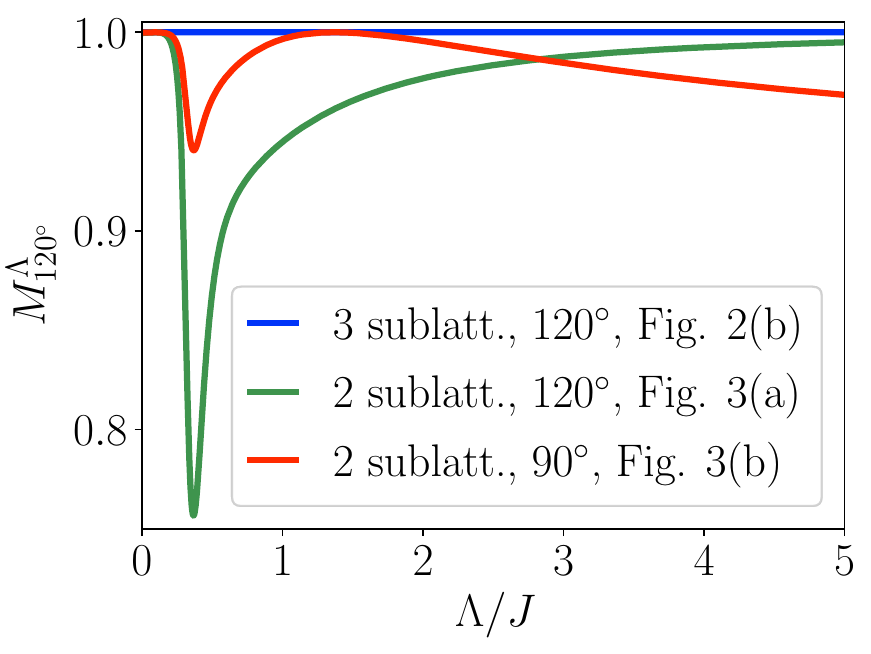}
        \put(-5,71){(c)}
        \end{overpic}
    \end{subfigure}
        \begin{subfigure}[t]{1\columnwidth}
        \centering
        \begin{overpic}[width = 0.9\textwidth]{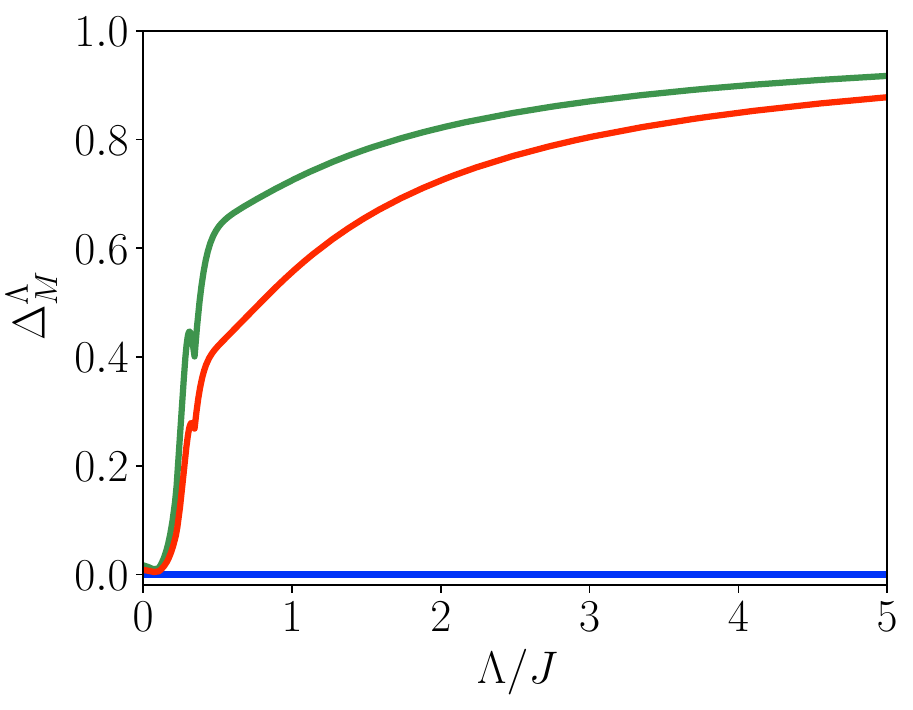}
        \put(-5,71){(d)}
        \put(44,16){\begin{overpic}[scale=.215]{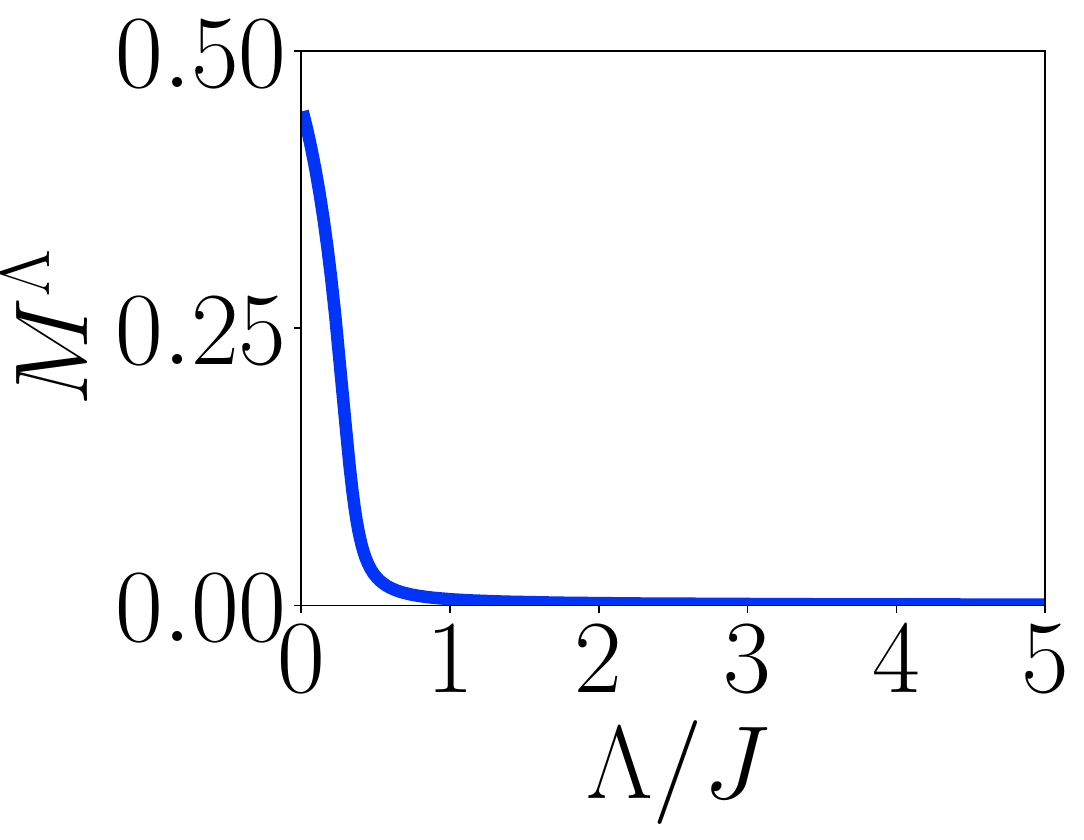}
        \end{overpic}}
        \end{overpic}
    \end{subfigure}
    \caption{(a), (b) Two choices of seed fields $\delta\hspace*{0.5pt}{\bm n}_i$ deviating from a $120\degree$ N\'eel pattern on the triangular lattice. Red dots indicate vanishing amplitudes $\delta$ on this sublattice. On the other two sublattices with identical $\delta$, the arrows depict the directions of ${\bm n}_i$ which either enclose (a) angles of $120\degree$ or (b) angles of $90\degree$. (c) $M_{120\degree}^\Lambda$ defined in Eq.~(\ref{eq:M120}) as a function of $\Lambda$ for the two seed fields of the triangular lattice Heisenberg antiferromagnet in (a) and (b) and for the perfect $120\degree$ seed field in Fig.~\ref{fig:staggeredMagnetization}(b). (d) $\Delta_M^\Lambda$ defined in Eq.~(\ref{eq:delta}) as a function of $\Lambda$ for the same seed fields as considered in (c). The inset in (d) shows the magnetization $M^\Lambda$ of the triangular lattice Heisenberg antiferromagnet as a function of $\Lambda$ for the ideal seed fields illustrated in Fig.~\ref{fig:staggeredMagnetization}(b).}
    \label{fig:TriangularOrderParameter}
\end{figure}

For all considered models, the magnetization converges at relatively small $L$, already around $L\approx 5$. At the largest considered correlation distances PFFRG finds magnetizations $M\approx0.436$ for the square lattice, $M\approx0.411$ for the honeycomb lattice and $M\approx0.442$ for the triangular lattice Heisenberg antiferromagnets. Compared to literature values ($M=0.3075(25)$~\cite{Runge92} from qMC for the square lattice, $M=0.22(3)$~\cite{Reger_1989} from qMC for the honeycomb lattice and $M=0.205(15)$~\cite{White07,huang23,Capriotti99} from variational Monte Carlo and DMRG for the triangular lattice) the PFFRG overestimates the magnetizations significantly, i.e., underestimates the impact of quantum fluctuations. However, we do not believe that these results indicate a general underestimation of quantum fluctuations within PFFRG since this has not been systematically observed in previous PFFRG applications for systems with TRS. For example, the extents of magnetically disordered phases compare well with other methods for the $J_1$-$J_2$ square lattice Heisenberg model~\cite{Reuther10}, the Heisenberg-Kitaev model~\cite{Reuther11}, the Shastry-Sutherland model~\cite{Keles22} or the non-Kramers nearest neighbor pyrochlore model~\cite{Lozano23}. Rather, we postulate that the overestimated magnetizations are a consequence of the specific conditions under which they are calculated. Accurate zero-field magnetizations require the correct buildup of magnetization when the renormalization group flow passes through the critical region at $\Lambda\approx\Lambda_c$ and small $\delta$. However, the critical region where the magnetization undergoes an abrupt increase and the susceptibility strongly peaks is prone to generating errors intrinsic to the method, as was already indicated by the discrepancies between the two approaches of calculating the susceptibility in Fig.~\ref{fig:PerturbedModels}(b) and (d). Specifically, close to criticality the vertex function flow is largely controlled by the RPA channel in our current truncation scheme. This flow through a critical RPA-dominated regime may impose a mean-field character (known to be generated in the RPA channel) on the magnetic correlations which may possibly explain the overestimation of the magnetizations down to $\Lambda\rightarrow0$.

While these observations indicate that zero-field magnetizations are problematic within PFFRG, we now demonstrate that the detection of the correct types of magnetic orders can be quite robust. For the magnetization of the triangular lattice Heisenberg model in Fig.~\ref{fig:staggeredMagnetization}(c), the seed fields $\delta\hspace*{0.5pt}{\bm n}_i$, imposed in the initial conditions, have been oriented according to the $120\degree$ N\'eel order in Fig.~\ref{fig:staggeredMagnetization}(b). However, the correct $120\degree$ spin configuration is even obtained if $\delta\hspace*{0.5pt}{\bm n}_i$ deviates substantially from this pattern. We demonstrate this for two choices of deviating seed fields shown in Fig.~\ref{fig:TriangularOrderParameter}(a) and (b). In both cases $\delta$ is non-vanishing on only two of the three sublattices of the $120\degree$ N\'eel order. These two sublattices have identical $\delta$ while ${\bm n}_i$ either enclose angles of $120\degree$ [see Fig.~\ref{fig:TriangularOrderParameter}(a)] or $90\degree$ [see Fig.~\ref{fig:TriangularOrderParameter}(b)].

We use two quantities to characterize the structure of the local magnetization that builds up during the renormalization group flow under these seed fields. The first, $M^\Lambda_{120\degree}$, probes the angles between the induced magnetizations ${\bm M}^\Lambda_{i\in \gamma}$ on the three sublattices $\gamma\in\{A,B,C\}$. It is defined by
\begin{equation}
M_{120\degree}^\Lambda=\frac{2}{3\sqrt{3}}|{\bm m}^\Lambda_A\times {\bm m}^\Lambda_B+{\bm m}^\Lambda_B\times {\bm m}^\Lambda_C+{\bm m}^\Lambda_C\times {\bm m}^\Lambda_A|,\label{eq:M120}
\end{equation}
where ${\bm m}_\gamma^\Lambda={\bm M}^\Lambda_{i\in\gamma}/|{\bm M}^\Lambda_{i\in\gamma}|$. The quantity $M^\Lambda_{120\degree}$ assumes the maximum value $M^\Lambda_{120\degree}=1$ only when ${\bm M}^\Lambda_{i\in \gamma}$ enclose the correct angles of $120\degree$ between each pair of spins (in which case they are coplanar). The second quantity, $\Delta_M^\Lambda$, probes differences in the magnitudes of the induced magnetizations ${\bm M}^\Lambda_{i\in \gamma}$ on the three sublattice, which in the ideal $120\degree$ N\'eel order are identical, $|{\bm M}^\Lambda_{i\in A}|=|{\bm M}^\Lambda_{i\in B}|=|{\bm M}^\Lambda_{i\in C}|$. We define $\Delta_M$ by
\begin{equation}
\Delta_M^\Lambda=\frac{M^\Lambda_\text{max}-M^\Lambda_\text{min}}{M^\Lambda_\text{max}}\label{eq:delta}
\end{equation}
where $M^\Lambda_\text{max}=\text{max}\{|{\bm M}^\Lambda_{i\in A}|,|{\bm M}^\Lambda_{i\in B}|,|{\bm M}^\Lambda_{i\in C}|\}$ is the maximum magnetization amplitude while $M^\Lambda_\text{min}=\text{min}\{|{\bm M}^\Lambda_{i\in A}|,|{\bm M}^\Lambda_{i\in B}|,|{\bm M}^\Lambda_{i\in C}|\}$ is the minimum magnetization amplitude. With these definitions the ideal $120\degree$ N\'eel order is characterized by $M^\Lambda_{120\degree}=1$ and $\Delta^\Lambda_M=0$.

Our results for $M^\Lambda_{120\degree}$ and $\Delta^\Lambda_M$ are shown in Fig.~\ref{fig:TriangularOrderParameter}(c) and (d), respectively. At large $\Lambda$ above criticality, $\Lambda>\Lambda_c\approx0.33J$, the reduction of $\Delta_M^\Lambda$ from its largest possible value $\Delta_M^\Lambda=1$ indicates the generation of a finite magnetization on the sublattice where the seed fields are zero. The criticality at $\Lambda=\Lambda_c$ acts as a disruption of the renormalization group flow where $M_{120\degree}^\Lambda$ deviates more strongly from the expected value $M_{120\degree}^\Lambda=1$. This resembles our previous observation of error generation in the critical region. However, for $\Lambda<\Lambda_c$ the renormalization group flow of $M_{120\degree}^\Lambda$ turns up again and $\Delta_M^\Lambda$ shows a sharp decrease such that in the limit $\Lambda\rightarrow0$ the precise values $M^\Lambda_{120\degree}=1$ and $\Delta^\Lambda_M=0$ are obtained. We additionally plot $M^\Lambda$ for ideal $120\degree$ seed fields in the inset of Fig.~\ref{fig:TriangularOrderParameter}(d) to indicate the $\Lambda$ region where the magnetization builds up.

A possible reason for the robustness of the $120\degree$ N\'eel order despite the observation of large errors in the magnetization amplitude $M^{\Lambda\rightarrow0}$ is that information on the spin pattern is not only contained in the magnetizations ${\bm M}^\Lambda_i$ (which follow from the self-energy) but also in the spin-spin correlations $\chi^\Lambda_{ij}$ (which follow from the two-particle vertex). The correct spin-spin correlations of the $120\degree$ N\'eel order in the antiferromagnetic triangular Heisenberg model have previously been obtained in an application of the PFFRG with TRS above the critical $\Lambda$ scale~\cite{Reuther11b}, where correlations are short-range. In other words, while an accurate result for $M^{\Lambda\rightarrow 0}$ requires the proper buildup of magnetization {\it in} the error-prone critical region, the correct spin configuration is already captured {\it above} $\Lambda_c$ and the criticality only induces an intermediate error in the magnetization pattern.

\begin{figure}[t]
        \centering
        \hspace{-4mm}
        \begin{overpic}[width = 0.5\textwidth]{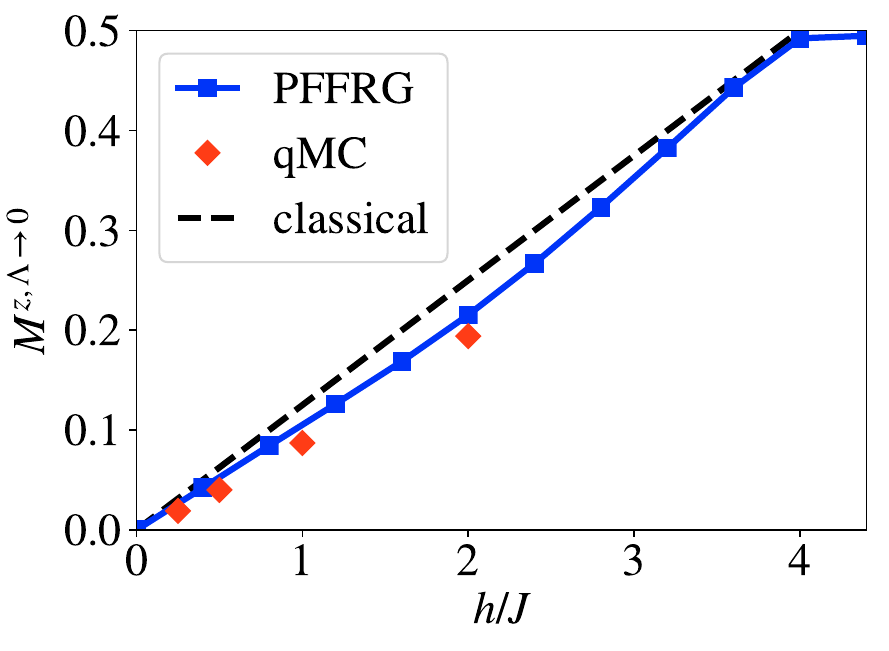}
        \put(66,20){\begin{overpic}[scale=.4]{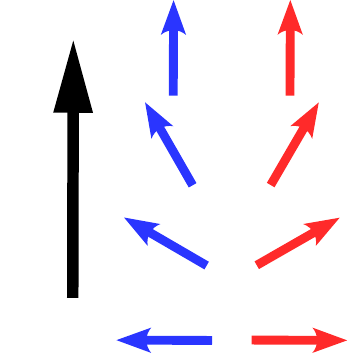}
        \put(0,42){\large $\bm{h}$}
        \end{overpic}}
        \end{overpic}
        \vspace{-6mm}
        \caption{Blue squares show the magnetization $M^{z,\Lambda\rightarrow 0}$ from PFFRG of the antiferromagnetic Heisenberg model on the square lattice as a function of the homogeneous field strength $h$, linearly extrapolated to zero seed fields, $\delta \rightarrow 0$. For comparison, red diamonds are qMC results from Ref.~\cite{Sandvik99}. The dashed black line is the linear classical magnetization curve. The inset illustrates the magnetization process schematically, where red and blue arrows depict spins on the two sublattices in a field ${\bm h}$ that increases from bottom to top.}
        \label{fig:SqMagnetizationCurve}
\end{figure}

\subsection{Magnetization process of the square lattice Heisenberg antiferromagnet}\label{sec:magnetization_curve}
As a last application of the PFFRG with magnetic fields, we calculate the magnetization curve of the square lattice Heisenberg antiferromagnet where (in contrast to the applications in Sec.~\ref{sec:flow_regularization} and~\ref{sec:zerofield}) a homogeneous magnetic field $h$ is considered, see Eq.~(\ref{eq:HeisenbergInBField}). This system has been studied in a variety of previous works~\cite{Yang97,Honecker04,Zhitomirsky98,Sandvik99}, including qMC~\cite{Sandvik99}. At $h\ll J$ the spins order antiferromagnetically in a direction perpendicular to the magnetic field, which, here, we apply along the $z$-direction. As $h$ increases, the spins continuously cant towards the $z$ axis, until they are ferromagnetically aligned at $h \geq 4J$ with a saturated longitudinal magnetization $M^{z}=1/2$. The inset in Fig.~\ref{fig:SqMagnetizationCurve} illustrates the magnetization process schematically.
A characteristic property of the magnetization curve is its non-linear behavior, where a small upward curvature results from a gradual suppression of zero-point fluctuations~\cite{Zhitomirsky98}. This is in contrast to the magnetization curve of the corresponding classical spin model which is strictly linear, $M^{z}_{\text{class}}=h/(8J)$, up to saturation.

To regularize PFFRG flows and to calculate the longitudinal magnetization $M^{z,\Lambda\rightarrow0}$, two fields need to be implemented, the homogeneous field $h$ and a staggered seed field ${\bm n}_i\perp (0,0,1)$ that generates the perpendicular antiferromagnetic spin order. The obtained magnetization curve, plotted in the main panel of Fig.~\ref{fig:SqMagnetizationCurve}, shows good qualitative agreement with qMC. Particularly, we find the expected upward curvature, which, however, is somewhat less pronounced than in qMC. This overestimation of $M^{z,\Lambda\rightarrow0}$ is possibly related to the overestimation of zero-field antiferromagnetic order on the square lattice in Sec.~\ref{sec:zerofield}. However, the errors in $M^{z,\Lambda\rightarrow 0}$ in Fig.~\ref{fig:SqMagnetizationCurve} are considerably smaller than those of the spontaneous magnetizations at $h=0$ in Sec.~\ref{sec:zerofield}. Specifically, the deviation of the magnetization from saturation, i.e., $1/2-M$ in the antiferromagnetic square lattice Heisenberg model at $h=0$ calculated with PFFRG is only $33\%$ of the exact qMC result for $1/2-M$. This can be compared to the extent to which PFFRG captures the deviations from the linear classical magnetization curve. In this context, PFFRG finds $71\%$ of the qMC result for $M^{z}_{\text{class}}-M^z$ at $h=2J$~\cite{Sandvik99}. The more accurate description of quantum fluctuations at $h=2J$ compared to $h=0$ is possibly a consequence of the fact that at $h=2J$ the PFFRG can already benefit from its perturbative error control at small $J/h$. More precisely, the PFFRG is known to be exact up to second perturbative order in the interaction strength $J$~\cite{Mueller23}.

We also note that PFFRG finds saturation $M^{z,\Lambda\rightarrow0}=1/2$ very accurately at the expected field strength $h=4J$ which may, similarly, be a consequence of the perturbative error control.

\section{Discussion}\label{sec:Discussion}
We have extended the PFFRG to treat spin Hamiltonians with finite magnetic fields and demonstrated its numerical feasibility. Our investigations of Heisenberg models on square, honeycomb and triangular lattices either in the presence of small magnetic seed fields generating finite magnetic order parameters or in the presence of homogeneous magnetic fields, provide important insights about the types of applications for which the method is best suited. If a physical observable does not require a renormalization group flow in $\Lambda$ through a critical regime close to a critical point we observe that it may be captured rather accurately. An example is the $120\degree$ N\'eel order on the triangular lattice which is correctly generated even if the seed fields have a very different structure. A possible explanation is that (short range) spin-spin correlations of magnetic ordering patterns are already included in the two-particle vertex above the critical $\Lambda$. On the other hand, if the precise value of a physical observable crucially depends on its generation at a critical $\Lambda$, it may be subject to large errors in the PFFRG. We observed this for the zero-field magnetizations of aniferromagnetic spin models which are significantly overestimated in the limit of vanishing seed fields. The renormalization group flows in the critical $\Lambda$ regions where the magnetizations build up, are dominated by the RPA channel of the two-particle vertex. When considered alone, this channel generates the order parameter susceptibility of a bare spin-mean field theory. This induces a mean-field bias in the critical $\Lambda$ region that may explain the overestimation of the amplitudes of magnetizations. In the future, it will be interesting to investigate whether multiloop PFFRG approaches~\cite{Kugler18_1,kugler18_2,Kiese22,Thoenniss20} can mitigate this problem. Such schemes include additional vertex contributions beyond mean-field, possibly reducing the mean-field bias near criticality.

Since our extended PFFRG has proven to be robust in detecting spin patterns, the investigation of complex magnetic orders which, e.g., often occur in magnetization plateaus represents a promising endeavor for future applications. However, the flow regularisation in such applications could require large computational resources since the numerical efforts increase with each spontaneously broken symmetry of a ground state. Therefore, ground states with large magnetic unit cells pose a challenge due to their broken translation symmetries and the resulting symmetry-inequivalent sites for each of which the flow equations have to be solved after applying the seed field. A further difficulty is that the PFFRG is intrinsically unable to find the spin configuration of a magnetization plateau in an unbiased manner. It can only detect those spin states whose broken symmetries are explicitly assumed by small seed fields.
On the other hand, our present studies show that the seed fields do not need to closely match the spin orientations in the actual ground state. We therefore still consider it worthwhile to investigate magnetization plateaus within the PFFRG, but we defer this to future work. The need to resolve complicated spin patterns via flow regularizations is eliminated when studying field-induced quantum spin liquids~\cite{Ivica21,banerjee18}, which represents another promising direction for future applications.

Finally, we mention that several code extensions might be important for obtaining accurate results in these future applications. This includes, e.g., a continuous frequency cutoff function, better resolution of characteristic features in the frequency dependence of vertex functions, consideration of their asymptotic frequency structure, or improved frequency integrations~\cite{Kiese22}. When applied to spin systems and observables, where the PFFRG can unfold its strengths, these methodological efforts could pay off and provide interesting insights into complex quantum phenomena.

\section*{Acknowledgements}
The authors thank Frederic Bippus, Nils Niggemann, Bj\"orn Sbierski and Benedikt Schneider for helpful discussions.
V.N. and J.R. gratefully acknowledge the computing time on the high-performance computing center Noctua 2 at the NHR Center PC2. This is funded by the Federal Ministry of Education and Research and the state governments participating on the basis of the resolutions of the GWK for the national high-performance computing at universities (www.nhr-verein.de/unsere-partner). The computations for this research were performed using computing resources under project hpc-prf-pm2frg.
Additionally, the authors would like to thank the HPC Service of ZEDAT and Tron cluster service at the Department of Physics, Freie Universität Berlin, for computing time.

\appendix

\section{\label{appendix:AnalyticSolution}Comparison of $T$ and $\Lambda$ for a free spin in a magnetic field}

In this appendix, we consider a free spin-$\frac{1}{2}$ in a magnetic field described by the Hamiltonian
\begin{equation}
\label{eq:freeSpin}
    \hat{\mathcal{H}} = -h \hat{S}^{z},
\end{equation}
and compare the analytical expressions for the magnetization and susceptibility when either the temperature $T$ or a sharp frequency cutoff $\Lambda$ act as a regulator. The calculation of these quantities does not require the solution of a renormalization group equation but only knowledge of the free Green function. The results will reveal qualitative differences between $T$ and $\Lambda$ and specify the expected flow behavior of interacting spin models at high magnetic fields.

\begin{figure}
        \centering
        \includegraphics[width = 0.4\textwidth]{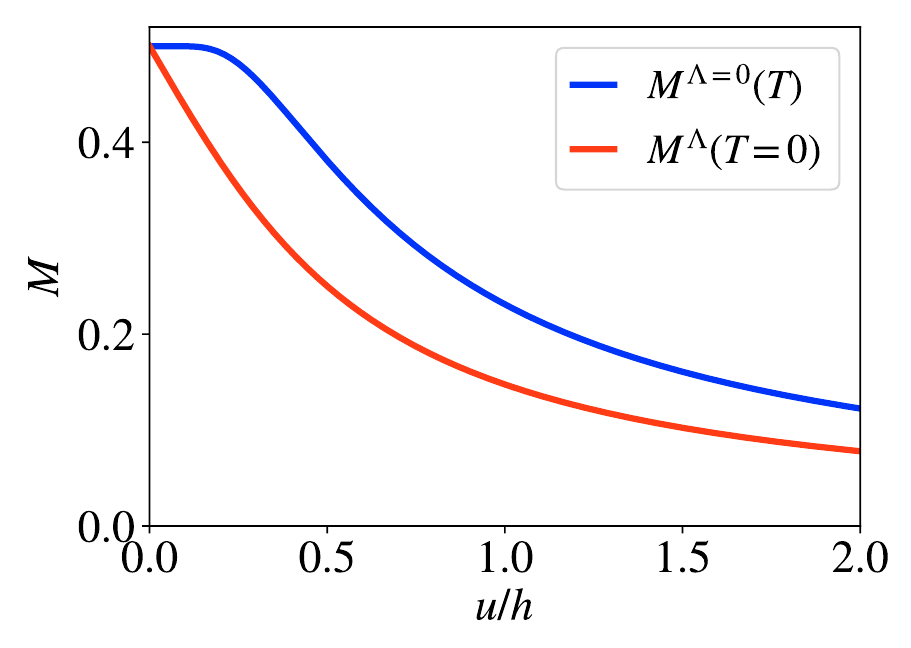}
        \caption{Magnetization $M$ of a free spin in a magnetic field $h$ either as a function of the temperature $T$ or as a function of the sharp frequency cutoff parameter $\Lambda$, both represented by the variable $u\in\{T,\Lambda\}$. The explicit functions are given in Eqs. \eqref{eq:analyticCutoffdependentMagnetization} and \eqref{eq:analyticTemperatureDependentMagnetization}.}
        \label{fig:nonInteractingM}
\end{figure}

The free pseudo-fermion Green function obtained from the Hamiltonian in \eqref{eq:freeSpin} is given by
\begin{align}
    &G^{\Lambda}(1'|1) = G^{\Lambda}_{0}(1'|1)\nonumber\\
    &=\theta(|\omega_{1}|-\Lambda)\delta(\omega_{1'}-\omega_{1})\left(i\omega_{1}\sigma^0+\frac{h}{2} \sigma^{z}\right)^{-1}_{\alpha_{1'}\alpha_{1}}.
\end{align}
Using the parameterization in Eq.~(\ref{eq:green_parameterization}) and inserting the Green function into Eq.~\eqref{eq:Magnetization} yields the $\Lambda$-dependent magnetization
\begin{align}
M^{\Lambda}(T=0) =& \frac{1}{2}- \frac{1}{\pi}\text{arctan}\left(\frac{2\Lambda}{h}\right).
            \label{eq:analyticCutoffdependentMagnetization}
\end{align}
In contrast, the temperature-dependent magnetization is given by
\begin{equation}
    M^{\Lambda=0}(T) = \frac{1}{2} \text{tanh}\left(\frac{h}{2T}\right).
    \label{eq:analyticTemperatureDependentMagnetization}
\end{equation}
Both functions are plotted in Fig.~\ref{fig:nonInteractingM} which demonstrates qualitative differences between $T$ and $\Lambda$. While the $T$-dependence of $M^{\Lambda=0}(T)$ has a vanishing slope at $T=0$, the $\Lambda$-dependence of $M^{\Lambda}(T=0)$ has a finite slope at $\Lambda=0$. Particularly, $T$ and $\Lambda$ are not related by the simple relation $\Lambda=2T/\pi$ that was proposed in Ref.~\cite{Iqbal2016} for $h=0$.

\begin{figure}
        \centering
        \includegraphics[width = 0.4\textwidth]{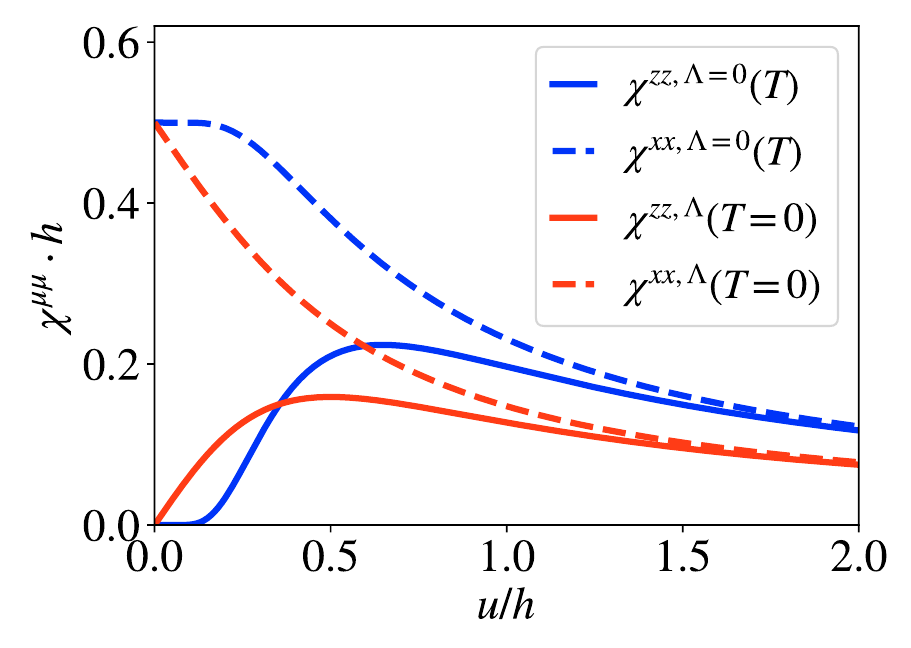}
        \caption{Susceptibilities $\chi^{\mu\mu}$ for $\mu=x,z$ of a free spin in a magnetic field $h$ along the $z$-axis. Shown is either the dependence on the temperature $T$ or on the sharp frequency cutoff $\Lambda$, both represented by the variable $u\in\{T,\Lambda\}$. The explicit functions are given by Eqs. \eqref{eq:chizzAnalytic}-\eqref{eq:chixxAnalytic2}.}
        \label{fig:nonInteractingChi}
\end{figure}

The corresponding susceptibilities $\chi^{\mu\mu}$ of the free spin are obtained by taking the derivative of the magnetization with respect to the external field, and are given by
\begin{align}
    \label{eq:chizzAnalytic}
    &\chi^{zz,\Lambda}(T=0) = \frac{1}{\pi h} \frac{1}{\frac{2\Lambda}{h}+\frac{h}{2\Lambda}},\\
\label{eq:chixxAnalytic}
    &\chi^{xx,\Lambda}(T=0) = \frac{1}{2h}\left[1 - \frac{2}{\pi} \text{arctan} \left(\frac{2\Lambda}{h}\right) \right],\\
        \label{eq:chizzAnalytic2}
    &\chi^{zz,\Lambda=0}(T) = \frac{1}{4T \text{cosh}\left(\frac{h}{2T}\right)^{2}},\\
\label{eq:chixxAnalytic2}
    &\chi^{xx,\Lambda=0}(T) = \frac{1}{2h} \text{tanh}\left(\frac{h}{2T}\right).
\end{align}
These functions are shown in Fig.~\ref{fig:nonInteractingChi} which reveals similar differences between $T$ and $\Lambda$ as in Fig.~\ref{fig:nonInteractingM}.

\section{\label{appendix:Meanfield}Mean-field magnetization as a function of $\Lambda$}

Here, we solve the self-consistent mean-field equation for the magnetization in a magnetic field $h$ where, in contrast to the usual formulation in terms of the temperature $T$, we use the sharp cutoff parameter $\Lambda$. When only keeping flow equation terms in Eqs.~\eqref{eq:SEFlowequationImplicit} and \eqref{eq:FlowEqComplete2} that contain an internal site summation $\sum_j$, the Hartree and RPA approximation is reproduced~\cite{Baez17}. 
In the absence of magnetic fields, the self-energy vanishes for any $\Lambda$. In this case, the RPA can be obtained by solving the two-particle vertex flow equation [Eq.~(\ref{eq:FlowEqComplete2})] only in the RPA channel, or, equivalently~\cite{kugler18_2}, by solving the Bethe-Salpeter self-consistency equation (the former is obtained from the latter by taking a derivative with respect to $\Lambda$). Finite magnetic fields enable the computation of the RPA susceptibility by taking the derivative of the Hartree magnetization with respect to the external magnetic field. In this approach, knowledge of the two particle-vertex is not needed.

For simplicity, we assume a nearest neighbor Heisenberg model with a site-independent self-energy, exposed to a uniform magnetic field along the $z$ axis, i.e. 
\begin{equation}
    \hat{\mathcal{H}} = J \sum_{\langle ij \rangle}  \hat{\bm{S}}_{i} \cdot \hat{\bm{S}}_{j} - h \sum_{i} \hat{S}^{z}_{i}.
\end{equation}
After the mapping to a pseudo-fermion Hamiltonian, the self-consistent equation for the Hartree self-energy~\cite{FetterWalecka} is given by
\begin{equation}
    \Sigma^{\Lambda}(1',1) = \sum_{2,2'} \frac{J_{||}(1',2'|1,2)}{4} G^{\Lambda}(2|2'),
\end{equation}
with
\begin{align}
J_{||}(1',2'|1,2) &= J_{i_1 i_2}\sum_{\mu} \sigma^{\mu}_{\alpha_{1'}\alpha_{1}} \sigma^{\mu}_{\alpha_{2'}\alpha_{2}}\times  \nonumber\\
    & \delta(\omega_{1'}+\omega_{2'}-\omega_{1}-\omega_{2}) \delta_{i_{1'}i_{1}}\delta_{i_{2'}i_{2}}.
\end{align}
Inserting the parameterizations of the self-energy and the Green function [see Eqs.~\eqref{eq:PropagatorPara3} and~\eqref{eq:green_parameterization}], one finds
\begin{equation}
    \gamma_{i}^{z,\Lambda} = \frac{1}{4\pi}\sum_{j} J_{ij} \int^{\infty}_{-\infty} d\omega g_{j}^{z,\Lambda}(\omega)
\end{equation}
Comparison with Eq. \eqref{eq:Magnetization} yields
\begin{equation}
   \gamma^{z,\Lambda} = \frac{cJM^{\Lambda}}{2},
\end{equation}
where $c$ is the coordination number of the lattice.
Inserting this relation into the Green function of Eq.~\eqref{eq:Magnetization} and evaluating the integral, leads to the self-consistent equation for the Hartree magnetization
\begin{equation}
\label{eq:MSelfconsistent}
    M^{\Lambda} = \frac{1}{2} \text{sgn}(h-cJM^{\Lambda}) - \frac{1}{\pi} \text{arctan}\left(\frac{2\Lambda}{h-cJM^{\Lambda}}\right).
\end{equation}
This equation allows to calculate the mean-field magnetizations in Fig.~\ref{fig:PerturbedModels}(a) and (c), dotted lines. For a free spin, $J=0$, the equation becomes identical to Eq.~\eqref{eq:analyticCutoffdependentMagnetization}. As mentioned before, at finite magnetic field $h$ the mean-field solutions formulated either in terms of $T$ or $\Lambda$ are qualitatively different. The corresponding self-consistent equation in $T$ involves a tanh-function instead of an arctan function.

\bibliography{references.bib}

\end{document}